\documentclass[aps,prd,twocolumn,superscriptaddress]{revtex4-2}

\usepackage{microtype}
\usepackage{graphicx}
\usepackage{dcolumn}
\usepackage{bm}
\usepackage{mathtools} 
\usepackage{tensor} 
\usepackage[hidelinks]{hyperref} 
\usepackage{amssymb}
\usepackage{mathtools}
\usepackage{mathrsfs}
\usepackage{amsmath} 
\usepackage{xcolor}
\usepackage{enumitem}
\usepackage{soul}
\usepackage{amsthm}

\usepackage{booktabs}

\usepackage[toc,page]{appendix}

\theoremstyle{definition}

\theoremstyle{remark}

\newcommand{\dd}{\mathrm{d}}
\newcommand{\epsz}{\epsilon_0}
\newcommand{\alp}{\alpha_0}

\newcommand{\R}{\mathbb R}
\newcommand{\Q}{\mathcal Q}
\newcommand{\C}{\mathcal C}

\newcommand{\Heis}{\mathfrak h_3}

\begin{document}

\title{Potential Space Symmetries in Ernst-like Formulations of Einstein-Maxwell/ModMax-Scalar field Theories}

\author{Leonel Bixano}
    \email{Contact author: leonel.delacruz@cinvestav.mx}
\author{Tonatiuh Matos}%
 \email{Contact author: tonatiuh.matos@cinvestav.mx}
\affiliation{Departamento de F\'{\i}sica, Centro de Investigaci\'on y de Estudios Avanzados del Instituto Politécnico Nacional, Av. Instituto Politécnico Nacional 2508, San Pedro Zacatenco, M\'exico 07360, CDMX.
}%

\date{\today}

\begin{abstract}
We complete the visible, hidden, sectorial, and discrete symmetries of Ernst-like potential spaces in stationary, axisymmetric Einstein–Maxwell–Scalar Field (EMSF) and Einstein–ModMax–Scalar Field (EMMSF) theories. In the real potential space
\(
    (f,\epsilon,\psi,\chi,\kappa),
\)
with target-space geometry defined by
\(
    d\epsilon-\psi\,d\chi ,
\)
we find the exact visible symmetries of the full target space and their solvable Lie algebra. We then characterize the hidden symmetries on specific invariant subspaces: the Ehlers transformation acts in the gravito-rotational sector, while electric and magnetic Harrison transformations act in the static electromagnetic subspaces. In the frozen EMMSF regime,
\(
    v=v_0,\quad w=w_0,
\)
we show how the EMSF sectorial transformations are deformed in the ModMax theory. We further show that the simultaneous coexistence of the electric and magnetic sectorial Harrison transformations imposes
\(
    \dd w=0
\)
and
\(
    \dd[(v^2+w^2)/w]=0
\),
thereby selecting precisely the frozen ModMax sector.

We study the Hamiltonian formulation of the potential space, the Noether charges, and the Casimir invariants of the sectorial algebras, relating these Casimirs to the invariant one-forms 
from previous works by the authors.
In purely harmonic branches of the \(A,B,C\) one-forms, the affine geodesic energy on the potential space is constant, so the quadrature for the metric function \(k\) is directly controlled by the Hamiltonian of the affine geodesic. The remaining functions \(\omega\) and \(A_\varphi\) follow from the Noether charges along the Killing directions of \(\epsilon\) and \(\chi\), and are expressed in terms of the dual harmonic function \(\widetilde{\mathfrak s}\).

We also examine the electric and magnetic Lewis-Weyl-Papapetrou frames and the discrete map between them. With a scalar field, this map acts as
\(
    \kappa\longmapsto \kappa^{-1},
    \quad
    \psi\longmapsto \chi,
    \quad
    \chi\longmapsto-\psi,
    \quad
    \epsilon\longmapsto\epsilon-\psi\chi .
\)
Finally, we apply the sectorial transformations to harmonic scalar-vacuum Weyl seeds with independent gravitational \(f\) and scalar \(\kappa\)  harmonics ~\cite{Herdeiro:2024oxn}. Frozen-ModMax electric and magnetic Harrison maps generate charged sectorial branches, while the Ehlers map generates the gravito-rotational branch. For these transformed solutions we provide the final quadratures for the physical functions \(k\), \(\omega\), and \(A_\varphi\).

\end{abstract}

\maketitle

\section{Introduction}
\label{sec:introduction}

Stationary and axisymmetric gravitational configurations exhibit a remarkable hidden structure. Upon dimensional reduction along the commuting Killing directions, the field equations can frequently be rewritten as a lower-dimensional gravity-coupled nonlinear sigma model, whose coordinates are gravitational, electromagnetic, and scalar potentials. In vacuum and electrovacuum gravity this observation is the basis of the classical Ernst formulation and its associated solution-generating techniques. Ernst's original formulation recasts the stationary axisymmetric Einstein and Einstein-Maxwell equations in terms of potentials defined on the reduced two-dimensional base \cite{Ernst:1967wx,Ernst:1967by}. Within this setting, Harrison transformations generate new Einstein-Maxwell configurations from known seeds \cite{Harrison:1968wue}, while Kinnersley organized the stationary Einstein-Maxwell system into a group-theoretic framework governed by its internal \(SU(2,1)\) structure \cite{Kinnersley:1977pg}. The Ehlers transformation, originally identified in the stationary vacuum sector, provides one of the simplest examples of a hidden transformation acting on the gravito-rotational sector, typically introducing twist or gravitomagnetic structure into an initially static configuration.

The inclusion of scalar fields modifies this picture in an essential way. In Einstein-Maxwell-Scalar Field theories the electric and magnetic sectors generally carry different scalar weights. Consequently, transformations that are global isometries of the Einstein-Maxwell target space may survive only on lower-dimensional invariant subspaces, or may require a nontrivial transformation of the scalar potential. This phenomenon is already visible in stationary Einstein-Maxwell-dilaton theory, where the Killing equations of the five-dimensional target space show that, for generic dilaton coupling, the symmetry algebra reduces to a maximal solvable subalgebra, whereas special critical couplings lead to enhanced hidden symmetry algebras  \cite{Galtsov:1995mb}. Related structures appear in Einstein-Maxwell-dilaton-axion systems, where Gal'tsov and collaborators found Ehlers-Harrison-type transformations, duality groups, and solution-generating maps in which the dilaton and axion participate nontrivially in the hidden symmetries \cite{Galtsov:1994pd,Galtsov:1995ssj,Galtsov:1996qko,Herrera-Aguilar:1998oct}. These works form part of the broader \emph{stringy gravity} integrability program, where stationary reductions are described by coset or Kahler sigma models and their associated finite and infinite-dimensional symmetry groups \cite{Galtsov:1994sjr}.

A persistent technical issue in this entire program is that generating the potentials is not the end of the construction. The remaining metric functions, in particular the Weyl-Lewis-Papapetrou function \(k\), must still be reconstructed from first-order quadratures. In explicit stationary axisymmetric solutions this step can be technically as important as the generation of the Ernst potentials themselves, as is clear in many exact electrovac constructions where the final metric is obtained only after the additional metric functions have been integrated or simplified into a usable closed form \cite{Manko:2019gvj}. One of the aims of the present work is to make this reconstruction transparent in harmonic branches: the coefficient entering the \(k\)-quadrature is identified with the Hamiltonian, or affine geodesic energy, of the corresponding target-space curve, while \(\omega\) and \(A_\varphi\) are fixed by the Noether charges associated with the Killing directions \(\epsilon,\chi\) of the potential space.

Recently, scalar-vacuum Weyl configurations have provided a useful arena in which to revisit this issue. Herdeiro showed that the
scalar-vacuum Weyl system admits two fully decoupled harmonic functions: one controlling the gravitational Weyl potential and another controlling the scalar sector. This leads to black holes or masses immersed in scalar multipolar universes and also admits a Kaluza-Klein oxidation to a five-dimensional generalized Weyl construction \cite{Herdeiro:2024oxn}. Independently, Bokulić and Herdeiro constructed generalized Harrison transformations for Einstein-ModMax theory, showing that sectorial electric and magnetic Harrison maps survive in the purely electric or purely magnetic sectors, they also considered the Einstein-dilaton-ModMax extension and applied these transformations to black holes and black diholes \cite{Bokulic:2025ucy}.

A complementary historical line, especially relevant for the present formulation, is the development of real potential-space methods for stationary and axisymmetric gravity with scalar and electromagnetic fields. In five-dimensional stationary axisymmetric gravity, Matos introduced one- and two-dimensional subspace methods both in spacetime and in potential space, recovering several known solutions as particular cases \cite{Matos:1994hm}. Matos and Plebański then described four-dimensional stationary axisymmetric vacuum gravity in terms of potential-space variables with target group \(SL(2,\mathbb R)\), leading to a classification of Weyl, Lewis, and related stationary classes \cite{Matos:1994qv}. In the dilatonic sector, Matos, Núñez and Quevedo constructed EMD solutions in terms of two arbitrary harmonic functions \cite{Matos:1995my}, while Matos and Mora obtained stationary EMD families with \(\alpha^2=3\), separating the gravitational and electromagnetic sectors through two independent generating functions \cite{Matos:1996km}. Later, Matos, Núñez and Ríos developed a functional-potential and Hamiltonian description for stationary axisymmetric EMD configurations \cite{Matos:2000za}, and related potential-space methods were extended to EMDA systems through subspaces of \(Sp(4,\mathbb R)\sim O(5)\) \cite{Matos:2009rp}. Importantly for the present work, Matos also applied this strategy to Einstein-Maxwell phantom fields, obtaining rotating and magnetized wormhole configurations \cite{Matos:2010pcd}.

The generalized Ernst potential spaces developed in Refs.~\cite{Bixano:2026xum,Bixano:2026ouq} continue this potential-space line by formulating the stationary axisymmetric problem directly in the real variables
\(
    (f,\epsilon,\psi,\chi,\kappa),
\)
with arbitrary dilatonic or phantom sign and, in the ModMax case, with a frozen nonlinear electromagnetic sector. This is the framework adopted in the present work. Its geometry is determined by the combination
\(
    d\epsilon-\psi\,d\chi ,
\)
which couples the gravitational twist potential to the electromagnetic potentials and fixes the structure of the associated target space.

Within this real potential-space formulation, we determine the visible symmetries of the full target space and organize the hidden transformations that survive only on invariant subspaces. The Ehlers map acts in the gravito-rotational sector, while the electric and magnetic Harrison maps act on the corresponding static electromagnetic subspaces. We also analyze the electric and magnetic Lewis-Weyl-Papapetrou frames and the discrete map relating them, in the presence of a dilaton or phantom scalar this map acts nontrivially on
\(
    \kappa,\quad \psi,\quad \chi,\quad \epsilon .
\)
We then extend the sectorial construction to the frozen Einstein-ModMax-Scalar Field regime,
\(
    v=v_0,
    \quad
    w=w_0,
\)
where the ModMax constitutive coefficients are taken to be constant. In this domain, the electric and magnetic Harrison maps undergo fixed ModMax renormalizations, yet they continue to act as explicit solution-generating maps within their respective sectorial subspaces and agree with the construction of \cite{Bokulic:2025ucy}.

We further analyze the Hamiltonian structure of the potential space, the Noether charges, and the Casimir invariants associated with these sectorial algebras. In the purely harmonic branches of the \(A,B,C\) one-forms, see \cite{Bixano:2026xum,Bixano:2026ouq}, the affine geodesic energy on the target space is constant. Consequently, the quadrature of the metric function \(k\) is governed by this Hamiltonian coefficient, while \(\omega\) and \(A_\varphi\) are determined by the Noether charges in the \(\epsilon\) and \(\chi\) directions and by the corresponding dual harmonic functions. Finally, we apply the frozen-ModMax electric and magnetic Harrison maps, together with the Ehlers map, to scalar-vacuum Weyl seeds with independent harmonic functions in \(f\) and \(\kappa\), of the type recently studied in Ref.~\cite{Herdeiro:2024oxn}. This combines scalar-vacuum Weyl seeds with sectorial frozen-ModMax solution-generating maps inside the real
generalized Ernst potential space.

The electric Lewis-Weyl-Papapetrou frame is written as
\begin{equation}\label{eq:Metrica_LEP-E}
    ds^2_{\rm e}
    =
    -f(dt-\omega d\varphi)^2
    +
    \dfrac{1}{f}
    \left[
        e^{2k}(d\rho^2+dz^2)
        +
        \rho^2 d\varphi^2
    \right],
\end{equation}
with electromagnetic potential \(A_{\rm e}=A_t(\rho,z)\,dt+A_\varphi(\rho,z)\,d\varphi,\) therefore the natural electromagnetic potential is
\begin{equation}\label{eq:4Potencial-E}
    \psi=2A_t .
\end{equation}

The magnetic frame \cite{Barrientos:2023dlf} is
\begin{equation}\label{eq:Metrica_LEP-M}
    ds^2_{\rm m}
    =
    f(d\varphi-\omega dt)^2
    +
    \dfrac{1}{f}
    \left[
        e^{2k}(d\rho^2+dz^2)
        -
        \rho^2 dt^2
    \right],
\end{equation}
where the natural electromagnetic potential is \(A_\varphi^{\rm m}\), and we write
\begin{equation}\label{eq:4Potencial-M}
    \psi_{\rm m}=2A_\varphi^{\rm m}.
\end{equation}


\section{Metric on the space of potentials and manifest symmetries} \label{Sec:EspacioPotenciales}
In this work, we adopt the potential-space metric that has been used in the analyses of Refs.~\cite{Bixano:2026xum, Matos:2000ai,Matos:2000za,Matos:2010pcd,Bixano:2026ouq}.

\subsection{Einstein-Maxwell-Scalar Field}

The potential space metric corresponding to the EMSF, is given by
{\footnotesize
\begin{equation}\label{eq:metric-EMS}
\dd s_{\rm M}^{2}
=
\frac{1}{2f^2}\left(\dd f^2+\Theta^2\right)
+
\frac{2\epsz}{\alp^2\kappa^2}\dd\kappa^2
-
\frac{\kappa^2}{2f}\dd\psi^2
-
\frac{1}{2f\kappa^2}\dd\chi^2 ,
\end{equation}
}
where we have defined \( \Theta:=\dd\epsilon-\psi\,\dd\chi\).

From the form of the metric \eqref{eq:metric-EMS}, we observe that \(\epsilon\) and \(\chi\) act as cyclic variables in the Lagrangian \eqref{eq:Lagrangiano-M} associated with \eqref{eq:metric-EMS}. Introducing the transformation \(\epsilon'=\epsilon+a\chi, \quad \psi'=\psi+a\) with $a\in \mathbb R$, the 1-form
{\small
\[
\Theta'=\dd\epsilon' - \psi'\dd\chi=(\dd\epsilon+a\dd\chi)- (\psi+a)\dd\chi=\dd\epsilon-\psi\dd\chi=\Theta
\]
}
remains unchanged. Therefore, the manifest symmetries are
\begin{subequations}
    \begin{equation}\label{eq:Simetria_visible-translations}
    \mathbf K_\epsilon=\partial_\epsilon,
    \qquad
    \mathbf K_\chi=\partial_\chi,
    \qquad
    \mathbf K_\psi=\partial_\psi+\chi\partial_\epsilon,
\end{equation}
\begin{equation}\label{eq:Simetria_Dg}
    \mathbf D_g=f\partial_f+\epsilon\partial_\epsilon
    +\frac{1}{2}\psi\partial_\psi
    +\frac{1}{2}\chi\partial_\chi,
\end{equation}
\begin{equation}\label{eq:Simetria_Ds}
    \mathbf D_s=\kappa\partial_\kappa+\chi\partial_\chi-\psi\partial_\psi.
\end{equation}
\end{subequations}
The symmetries \eqref{eq:Simetria_visible-translations} are associated, respectively, with the translation of the twist (\( \mathbf K_\epsilon \)), the magnetic translation (\( \mathbf K_\chi \)), and the compensated electric shift (\( \mathbf K_\psi \)); meanwhile, \eqref{eq:Simetria_Dg} describes the gravitational dilation and \eqref{eq:Simetria_Ds} the scalar-electromagnetic dilation.

Therefore, the mappings of the manifest symmetries are
{\footnotesize
\begin{subequations}\label{eq:Mapeos-Visible}
    For \(\mathbf K_\epsilon\):
    \begin{equation}
        \epsilon'=\epsilon+a,
        \qquad
        f'=f,
        \qquad
        \psi'=\psi,
        \qquad
        \chi'=\chi,
        \qquad
        \kappa'=\kappa.
    \end{equation}
    For \(\mathbf K_\chi\):
    \begin{equation}
        \chi'=\chi+p,
        \qquad
        f'=f,
        \qquad
        \epsilon'=\epsilon,
        \qquad
        \psi'=\psi,
        \qquad
        \kappa'=\kappa.
    \end{equation}
    For \(\mathbf K_\psi\):
    \begin{equation}
        \psi'=\psi+q,
        \qquad
        \epsilon'=\epsilon+q\chi,
        \qquad
        f'=f,
        \qquad
        \chi'=\chi,
        \qquad
        \kappa'=\kappa.
    \end{equation}
    For \(\mathbf D_g\):
    \begin{equation}
        f'=e^\lambda f,
        \quad
        \epsilon'=e^\lambda\epsilon,
        \quad
        \psi'=e^{\lambda/2}\psi,
        \quad
        \chi'=e^{\lambda/2}\chi,
        \quad
        \kappa'=\kappa.
    \end{equation}
    For \(\mathbf  D_s\):
    \begin{equation}
        \kappa'=e^\mu\kappa,
        \qquad
        \chi'=e^\mu\chi,
        \qquad
        \psi'=e^{-\mu}\psi,
        \qquad
        f'=f,
        \qquad
        \epsilon'=\epsilon,
    \end{equation}
\end{subequations}
}
with \( \{ a,p,q,\lambda,\mu\}\in \mathbb{R} \).

Regarding the algebra of the manifest symmetries, and adopting the convention \([X,Y]=XY-YX\), the commutators that do not vanish are
{\footnotesize
\begin{equation*}
    \begin{aligned}
        &[\mathbf K_\chi,\mathbf K_\psi]=\mathbf K_\epsilon,  \\
        &[\mathbf D_g,\mathbf K_\epsilon]=-\mathbf K_\epsilon,
        \quad
        [\mathbf D_g,\mathbf K_\chi]=-\frac{1}{2}\mathbf K_\chi,
        \quad
        [\mathbf D_g,\mathbf K_\psi]=-\frac{1}{2}\mathbf K_\psi,
        \\
        &[\mathbf D_s,\mathbf K_\chi]=-\mathbf K_\chi,
        \qquad
        [\mathbf D_s,\mathbf K_\psi]=\mathbf K_\psi,
        \qquad
        [\mathbf D_s,\mathbf K_\epsilon]=0.
    \end{aligned}
\end{equation*}
}
Therefore, these generators are isomorphic to the Heisenberg algebra.
\begin{equation}
    \langle \mathbf K_\epsilon,\mathbf K_\chi,\mathbf K_\psi\rangle\simeq\Heis,
\end{equation}
and the complete manifest algebra is
\begin{equation}
    \mathfrak g_{\rm vis}
    =
    \Heis\rtimes\left(\R_g\oplus\R_s\right).
\end{equation}

\subsection{Einstein-ModMax-Scalar Field}\label{SubSec:Einstein-ModMax-Scalar Field}
For the EMMSF, the metric of the potential space is 
{\footnotesize
\begin{equation}\label{eq:metric-EMMS}
    \dd s_{\rm MM}^{2}
    =
    \frac{1}{2f^2}\left[\dd f^2+\Theta^2\right]
    +
    \frac{2\epsz}{\alp^2\kappa^2}\dd\kappa^2
    -
    \frac{1}{2f}\left[
        w\kappa^2\dd\psi^2
        +\frac{\left(\dd\chi+v\kappa^2\dd\psi\right)^2}{w\kappa^2}
    \right].
\end{equation}
}

In general, \(v,w\) are functions that depend on \((\kappa,k,D\psi,\tilde{D}\psi,D\chi,\tilde{D}\chi;\gamma_0)\), where \(\gamma_0\) is the ModMax deformation parameter, and $D:=(\partial_\rho,\partial_z)^T$, $\Tilde{D}=(\partial_z,-\partial_\rho)^T$,are the operators defined in \cite{Bixano:2026xum,Bixano:2026ouq}.

The only block which differs from the EMSF case is the electromagnetic one,
\[
    -
    \frac{1}{2f}
    \left[
        w\kappa^2\dd\psi^2
        +
        \frac{
        \left(\dd\chi+v\kappa^2\dd\psi\right)^2
        }{
        w\kappa^2
        }
    \right].
\]

The Maxwell-scalar block is recovered for \(v=0, w=1\). 

We now show precisely when the manifest EMSF symmetries remain symmetries of the ModMax target. Let 
\[ X \in \{
\mathbf K_\epsilon,\quad
\mathbf K_\chi,\quad
\mathbf K_\psi,\quad
\mathbf D_g,\quad
\mathbf D_s \},
\]
the ModMax block is invariant under the corresponding EMSF transformation if and only if the transformation also preserves the ModMax coefficients, i.e.
\begin{equation}\label{eq:MM-visible-condition}
        X\in \mathfrak g_{\rm vis}
        \;:\;
        X(v)=0,\quad X(w)=0
\end{equation}
Indeed, the potentials \(\psi, \chi\) act as cyclic variables of the geodesic Lagrangian \eqref{eq:Lagrangiano-MM} corresponding to the metric \eqref{eq:metric-EMMS}. As a result, \(\mathbf K_\epsilon=\partial_\epsilon\) and \(\mathbf K_\chi=\partial_\chi\) represent Killing vectors precisely when the conditions
\[
\partial_\epsilon v=0,
\quad
\partial_\epsilon w=0,
\quad
\partial_\chi v=0,
\quad
\partial_\chi w=0,
\]
are fulfilled. Conversely, if
\[
    \left(\partial_\psi+\chi\partial_\epsilon\right)v=0,
    \qquad
    \left(\partial_\psi+\chi\partial_\epsilon\right)w=0,
\]
hold, then \(\mathbf K_\psi\) is likewise a Killing vector in the ModMax frame.

The two nontrivial checks are the two dilations. For the gravitational dilation, we have
\(
\dd f'=e^\lambda\dd f,
\quad
\Theta'=e^\lambda\Theta,
\)
therefore
{\footnotesize
\[
    \frac{1}{2f'^2}
    \left(
        \dd f'^2+\Theta'^2
    \right)
    =
    \frac{1}{2e^{2\lambda}f^2}
    e^{2\lambda}
    \left(
        \dd f^2+\Theta^2
    \right)
    =
    \frac{1}{2f^2}
    \left(
        \dd f^2+\Theta^2
    \right).
\]
}
The scalar block remains the same since \(\kappa'=\kappa\) according to \eqref{eq:Mapeos-Visible}. For the ModMax electromagnetic sector, we have
\(
\dd\psi'=e^{\lambda/2}\dd\psi,
\quad
\dd\chi'=e^{\lambda/2}\dd\chi,
\)
which implies \( w\kappa^2\dd\psi'^2 = e^\lambda w\kappa^2\dd\psi^2 \), and
\(
\dd\chi' + v\kappa^2\dd\psi' = e^{\lambda/2}\left(\dd\chi + v\kappa^2\dd\psi\right).
\)
Hence
\[
    \frac{
    \left(
        \dd\chi'
        +
        v\kappa^2\dd\psi'
    \right)^2
    }{
    w\kappa^2
    }
    =
    e^\lambda
    \frac{
    \left(
        \dd\chi
        +
        v\kappa^2\dd\psi
    \right)^2
    }{
    w\kappa^2
    }.
\]
The full electromagnetic bracket therefore scales as \(e^\lambda\), while the prefactor scales as
\(
-\frac{1}{2f'}
=
-\frac{1}{2e^\lambda f}.
\)
The two factors cancel. Thus \(\mathbf D_g\), from \eqref{eq:Simetria_Dg} is a ModMax Killing vector provided
\[
    \mathbf D_g(v)=0,
    \qquad
    \mathbf D_g(w)=0.
\]

For the scalar-electromagnetic dilation, we find
\(
\Theta'
=
\dd\epsilon'
-
\psi'\dd\chi'
=
\dd\epsilon
-
(e^{-\mu}\psi)(e^\mu\dd\chi)
=
\Theta,
\)
and likewise
\(
\frac{\dd\kappa'^2}{\kappa'^2}
=
\frac{e^{2\mu}\dd\kappa^2}{e^{2\mu}\kappa^2}
=
\frac{\dd\kappa^2}{\kappa^2}.
\)
For the ModMax electromagnetic sector, one similarly obtains
\(
\kappa'^2\dd\psi'^2
=
(e^{2\mu}\kappa^2)(e^{-2\mu}\dd\psi^2)
=
\kappa^2\dd\psi^2,
\)
and \( 
    \dd\chi'
    +
    v\kappa'^2\dd\psi'
    =
    e^\mu\dd\chi
    +
    v(e^{2\mu}\kappa^2)(e^{-\mu}\dd\psi)
    =
    e^\mu
    \left(
        \dd\chi+v\kappa^2\dd\psi
    \right).
\)
Therefore, 
\[ 
    \frac{
    \left(
        \dd\chi'
        +
        v\kappa'^2\dd\psi'
    \right)^2
    }{
    w\kappa'^2
    }
    =
    \frac{
    e^{2\mu}
    \left(
        \dd\chi+v\kappa^2\dd\psi
    \right)^2
    }{
    we^{2\mu}\kappa^2
    }
    =
    \frac{
    \left(
        \dd\chi+v\kappa^2\dd\psi
    \right)^2
    }{
    w\kappa^2
    } 
\]

Thus \(\mathbf D_s\), from \eqref{eq:Simetria_Ds} is a ModMax Killing vector provided
\[
    \mathbf D_s(v)=0,
    \qquad
    \mathbf D_s(w)=0.
\]

Hence the visible ModMax algebra is
\begin{equation}
    \mathfrak g_{\rm vis}^{\rm MM}
    =
    \left\{
        X\in\mathfrak g_{\rm vis}
        \;:\;
        X(v)=0,\quad X(w)=0
    \right\}.
\end{equation}
In particular, in the frozen sector
\(
v=v_0,
\quad
w=w_0,
\)
the conditions \(X(v)=X(w)=0\) hold automatically, and the full EMSF manifest algebra survives, thus 
\( \mathfrak g_{\rm vis}^{\rm MM,0}
    =
    \Heis\rtimes
    \left(
        \R_g\oplus\R_s
    \right).\)

\section{Hidden Symmetries}
\label{sec:Hidden Symmetries}
To obtain the hidden symmetries we will use the Killing equation \(\nabla_{(a}K_{b)}=0\), or equivalently in coordinates
\begin{equation}\label{eq:Ecuacion De Killing}
    (\mathfrak L_K g)_{ab}=K^c\partial_c g_{ab}+g_{cb}\partial_a K^c+g_{ac}\partial_b K^c=0.
\end{equation}
\subsection{Ehlers transformation}
Let \( \psi=0, \quad \chi=0 \). Then the metric \eqref{eq:metric-EMS} takes the form 
\[
    \dd s_{\rm grav}^{2}
    =
    \frac{\dd f^2+\dd\epsilon^2}{2f^2}
    +
    \frac{2\epsz}{\alp^2\kappa^2}\dd\kappa^2,
\]
so that \( g_{ff}=g_{\epsilon\epsilon}=1/(2f^2) \) and \(g_{f\epsilon}=0\). Selecting \(K=A(f,\epsilon)\partial_f+B(f,\epsilon)\partial_\epsilon\), the Killing equation \eqref{eq:Ecuacion De Killing} yields
\( f \partial_f A-A=0 \), \(\partial_\epsilon B=A/f\), and \(\partial_f B+\partial_\epsilon A=0\). From the first equation we obtain \(A=fF(\epsilon)\). Substituting into the second gives \(\partial_\epsilon B=F(\epsilon)\), while the third becomes \(\partial_f B=-fF'(\epsilon)\). Requiring integrability (\(\partial_\epsilon\partial_f B=\partial_f \partial_\epsilon B\)) leads to \(F''(\epsilon)=0\), hence \(F(\epsilon)=c_1+2c_2\epsilon\). Consequently, \(A=f(c_1+2c_2\epsilon)\), and integrating for \(B\) gives \(B=c_0+c_1\epsilon+c_2(\epsilon^2-f^2)\). Therefore, the most general Killing vector in this sector is
\begin{equation}\label{eq:Killing-General-Elhers}
    K
    =
    f(c_1+2c_2\epsilon)\partial_f
    +
    \left[
        c_0+c_1\epsilon+c_2(\epsilon^2-f^2)
    \right]\partial_\epsilon .
\end{equation}
Equivalently,
\(K=c_0\mathbf T_E+c_1\mathbf D_E+c_2\mathbf E\), with
\begin{equation}\label{eq:Generadores-Ehlers-M}
    \mathbf T_E=\partial_\epsilon,
    \quad
    \mathbf D_E=f\partial_f+\epsilon\partial_\epsilon,
    \quad
    \mathbf E=2\epsilon f\partial_f+(\epsilon^2-f^2)\partial_\epsilon.
\end{equation}
The constants \(c_0,c_1,c_2\) only select a basis of the three-dimensional Killing algebra: \(c_0=1\) gives the shift \(\mathbf T_E\), \(c_1=1\) gives the scale generator \(\mathbf D_E\), and \(c_2=1\) gives the nonlinear Ehlers generator \(\mathbf E\). 

Its commutators are
{\small
\begin{equation}\label{eq:Conmutadores-Ehlers-M}
    [\mathbf D_E,\mathbf T_E]=-\mathbf T_E,
    \qquad
    [\mathbf D_E,\mathbf E]=\mathbf E,
    \qquad
    [\mathbf T_E,\mathbf E]=2 \mathbf D_E,
\end{equation}
}
so that
\[
\langle \mathbf T_E,\mathbf D_E,\mathbf E\rangle\cong \mathfrak{sl}(2,\R).
\]

The Ehlers transformation is derived by performing the integration using \(\tau = \epsilon + i f\). Because \(\mathbf E(\tau) = \tau^2\), the relation \(\frac{\dd \tau}{\dd s} = \tau^2\) is satisfied, and thus the associated Ehlers maps are
\begin{equation}\label{eq:Mapeos-Ehlers-M}
    \begin{aligned}
        f'&=\frac{f}{(1-s\epsilon)^2+s^2f^2}, \qquad \kappa'=\kappa\\
        \epsilon'&=
        \frac{\epsilon-s(\epsilon^2+f^2)}{(1-s\epsilon)^2+s^2f^2},
    \end{aligned}
\end{equation}
where \(s \in \mathbb{R}\).

\subsection{Magnetic Harrison transformation}\label{SubSec:Magnetic Harrison transformation}
Taking into account the magnetic static subspace \(\epsilon=0,\quad\psi=0\), the target-space metric \eqref{eq:metric-EMS} reduces to
\begin{equation}\label{eq:Magnetic-Subspace-Metric}
    \dd s_{\rm mag}^{2}
    =
    \frac{\dd f^2}{2f^2}
    +
    \frac{2\epsilon_0}{\alpha_0^2\kappa^2}\dd\kappa^2
    -
    \frac{\dd\chi^2}{2f\kappa^2}.
\end{equation}
We use the coordinates \(x^a=(f,\chi,\kappa),\) so that the non-vanishing metric components are \( g_{ff}=\frac{1}{2f^2}, \quad g_{\chi\chi}=-\frac{1}{2f\kappa^2}, \quad g_{\kappa\kappa}= \frac{2\epsilon_0}{\alpha_0^2\kappa^2}\) .

Let us take a general Killing vector field on this three-dimensional magnetic subspace, \( K = K^f(f,\chi,\kappa)\partial_f + K^\chi(f,\chi,\kappa)\partial_\chi + K^\kappa(f,\chi,\kappa)\partial_\kappa\). The Killing equation to be solved is \eqref{eq:Ecuacion De Killing}, therefore, the \((ff)\) component yields
\[
        K^f\partial_f g_{ff}
        +
        2g_{ff}\partial_fK^f
        =
        0.
\]
Since \(\partial_f g_{ff} = -\frac{1}{f^3},\) one obtains \( -\frac{K^f}{f^3} + \frac{1}{f^2}\partial_fK^f = 0,\) or equivalently \(f\partial_fK^f-K^f=0\). Therefore, \( K^f=fA(\chi,\kappa)\).
Analogously, the \((\kappa\kappa)\) component gives
\[
    K^\kappa\partial_\kappa g_{\kappa\kappa}
    +
    2g_{\kappa\kappa}\partial_\kappa K^\kappa
    =
    0.
\]

Using \(\partial_\kappa g_{\kappa\kappa}  =  -\frac{4\epsilon_0}{\alpha_0^2\kappa^3}\), we get \(\kappa\partial_\kappa K^\kappa-K^\kappa=0,\)  and hence \(K^\kappa=\kappa C(f,\chi)\).

The mixed component \((f\chi)\) gives
\[
    g_{\chi\chi}\partial_fK^\chi
    +
    g_{ff}\partial_\chi K^f
    =
    0,
\]
equivalently \(-\frac{1}{2f\kappa^2}\partial_fK^\chi  +  \frac{1}{2f^2}\partial_\chi K^f  =  0\).  Substituting \(K^f=fA(\chi,\kappa)\), this becomes \(\partial_fK^\chi = \kappa^2\partial_\chi A\). Therefore \(K^\chi  =  f\kappa^2\partial_\chi A(\chi,\kappa)  +  B(\chi,\kappa)\).

The mixed component \((f\kappa)\) gives
\[
    g_{\kappa\kappa}\partial_fK^\kappa
    +
    g_{ff}\partial_\kappa K^f
    =
    0,
\]
or \(\partial_\kappa A + \frac{4\epsilon_0 f}{\alpha_0^2\kappa}\partial_f C  =  0\). Imposing the remaining Killing equations rules out any logarithmic contributions, and thus we arrive at \( A=A(\chi), \qquad C=C(\chi).\)

The \((\chi\chi)\) component gives
\[
    K^f\partial_f g_{\chi\chi}
    +
    K^\kappa\partial_\kappa g_{\chi\chi}
    +
    2g_{\chi\chi}\partial_\chi K^\chi
    =
    0.
\]
Using \(\partial_f g_{\chi\chi}  =  \frac{1}{2f^2\kappa^2},  \quad  \partial_\kappa g_{\chi\chi} =  \frac{1}{f\kappa^3}\), this equation reduces to \(\partial_\chi K^\chi =  \frac12 A+C\). Substituting \( K^\chi=f\kappa^2A'(\chi)+B(\chi,\kappa)\), one gets \(f\kappa^2A''(\chi) + \partial_\chi B = \frac12A+C\). Since the right-hand side does not depend on \(f\), it follows that \(A''(\chi)=0\). Hence \(A(\chi)=a\chi+b\), where \(a,b\) are constant values.

Finally, the \((\chi\kappa)\) component gives
\[
    g_{\kappa\kappa}\partial_\chi K^\kappa
    +
    g_{\chi\chi}\partial_\kappa K^\chi
    =
    0.
\]
Equivalently, \(\frac{2\epsilon_0}{\alpha_0^2\kappa^2} \partial_\chi K^\kappa - \frac{1}{2f\kappa^2} \partial_\kappa K^\chi = 0\). Since \(K^\kappa=\kappa C(\chi),  \quad  K^\chi=af\kappa^2+B(\chi,\kappa)\), one obtains \(\partial_\kappa B=0\), and \(C'(\chi)=\frac{\alpha_0^2}{2\epsilon_0}a\). Therefore \(C(\chi)=\frac{\alpha_0^2}{2\epsilon_0}a\chi+c\),  where \(c\) is a constant. Moreover, \(\partial_\chi B  = \frac12(a\chi+b) + \frac{\alpha_0^2}{2\epsilon_0}a\chi+c\). Thus \(B(\chi) = \frac{\alpha_0^2+\epsilon_0}{4\epsilon_0}a\chi^2 + \left(  \frac{b}{2}+c \right)\chi + d\),  where \(d\) is another integration constant.

Therefore, the most general Killing vector on the magnetic sector is
\begin{equation}\label{eq:General-Killing-Magnetic-Sector}
\begin{aligned}
K_M
={}&
f(a\chi+b)\partial_f
\\
&+
\left[
a
\left(
f\kappa^2+
\frac{\alpha_0^2+\epsilon_0}{4\epsilon_0}\chi^2
\right)
+
\left(
\frac{b}{2}+c
\right)\chi
+d
\right]\partial_\chi
\\
&+
\kappa
\left(
\frac{\alpha_0^2}{2\epsilon_0}a\chi+c
\right)\partial_\kappa .
\end{aligned}
\end{equation}
The constants \(a,b,c,d\) indicate that the magnetic sector possesses four independent Killing directions. A natural basis can be constructed by isolating the coefficients corresponding to each of these four constants. Nonetheless, the Harrison subalgebra does not coincide with the full four-dimensional Killing algebra. Instead, it is the three-dimensional subalgebra generated by the translation \(\mathbf P_M\), the nonlinear charging generator \(\mathbf H_M\), and their commutator \(\mathbf J_M = [\mathbf P_M,\mathbf H_M]\).

Indeed, choosing \(d=1\), with \(a=b=c=0\), gives \(\mathbf P_M=\partial_\chi\).

Choosing \(a=1\), with \(b=c=d=0\), gives the magnetic Harrison generator \(\mathbf H_M =  f\chi\partial_f+  \left(  f\kappa^2+  \frac{\alpha_0^2+\epsilon_0}{4\epsilon_0}\chi^2  \right)\partial_\chi+ \frac{\alpha_0^2}{2\epsilon_0}\kappa\chi\partial_\kappa\). 

Then the commutator \([\mathbf P_M,\mathbf H_M] = \partial_\chi \mathbf H_M\), gives \([\mathbf P_M,\mathbf H_M]  =  f\partial_f  +  \frac{\alpha_0^2+\epsilon_0}{2\epsilon_0}\chi\partial_\chi  +  \frac{\alpha_0^2}{2\epsilon_0}\kappa\partial_\kappa\). Therefore we identify \(\mathbf J_M  =  f\partial_f  +  \frac{\alpha_0^2+\epsilon_0}{2\epsilon_0}\chi\partial_\chi  +  \frac{\alpha_0^2}{2\epsilon_0}\kappa\partial_\kappa\).

Where we have defined 
\begin{equation}\label{eq:beta-gamma-magnetic}
    \beta:=\frac{\alpha_0^2}{2\epsilon_0},
    \qquad
    \gamma:=\frac{\alpha_0^2+\epsilon_0}{4\epsilon_0},
\end{equation}
the magnetic Harrison generators can be written compactly as
\begin{subequations}\label{eq:Generadores-HarrisonMagnetico-M}
    \begin{equation}
        \mathbf P_M=\partial_\chi,
    \end{equation}
    \begin{equation}
        \mathbf J_M=f\partial_f+2\gamma\chi\partial_\chi+\beta\kappa\partial_\kappa,
    \end{equation}
    \begin{equation}
        \mathbf H_M=f\chi\partial_f+
        (f\kappa^2+\gamma\chi^2)\partial_\chi+
        \beta\kappa\chi\partial_\kappa.
    \end{equation}
\end{subequations}

The remaining independent Killing direction is obtained by choosing \(c=1\), with \(a=b=d=0\). This gives \(\mathbf S_M=\chi\partial_\chi+\kappa\partial_\kappa\), this is a sectorial scaling symmetry. It is a genuine Killing vector of the magnetic metric, because the combinations \(\frac{\dd\chi^2}{\kappa^2}, \quad \frac{\dd\kappa^2}{\kappa^2}\) are invariant under \( \chi\mapsto e^s\chi, \quad \kappa\mapsto e^s\kappa, \quad f\mapsto f\). 

Nevertheless, \(\mathbf S_M\) is not the Harrison generator, the Harrison generator is \(\mathbf H_M\), because it is the nonlinear Killing direction that generates a non-trivial magnetic potential from a seed with \(\chi=0\), i.e. \( \left.\mathbf H_M^\chi\right|_{\chi=0}  =  f\kappa^2\). 

The commutators of the Harrison sector are
\begin{equation}\label{eq:Conmutadores-HarrisonMagnetico-M}
\begin{aligned}
    [\mathbf J_M,\mathbf P_M]
    &=
    -2\gamma \mathbf P_M,
    \\
    [\mathbf J_M,\mathbf H_M]
    &=
    2\gamma \mathbf H_M,
    \\
    [\mathbf P_M,\mathbf H_M]
    &=
    \mathbf J_M.
\end{aligned}
\end{equation}
Therefore, when \(\gamma\neq0 \),  the algebra generated by \(\mathbf P_M,\mathbf J_M,\mathbf H_M\) is isomorphic to \(\mathfrak{sl}(2,\mathbb R)\) after a suitable rescaling. Explicitly, one may define
\[
    e=\mathbf P_M,
    \qquad
    h=-\frac{1}{\gamma}\mathbf J_M,
    \qquad
    f_{\mathfrak{sl}}=-\frac{1}{\gamma}\mathbf H_M,
\]
which satisfy
\[
    [h,e]=2e,
    \qquad
    [h,f_{\mathfrak{sl}}]=-2f_{\mathfrak{sl}},
    \qquad
    [e,f_{\mathfrak{sl}}]=h.
\]

The complete four-dimensional Killing algebra of the magnetic sector is larger than the Harrison subalgebra. The additional Killing vector \(\mathbf S_M\) satisfies
{\footnotesize
\[
    [\mathbf S_M,\mathbf P_M]=-\mathbf P_M,
    \qquad
    [\mathbf S_M,\mathbf H_M]=\mathbf H_M,
    \qquad
    [\mathbf S_M,\mathbf J_M]=0.
\]
}
For \(\gamma\neq0\), the combination \( \mathbf Z_M = \mathbf S_M-\frac{1}{2\gamma}\mathbf J_M \) commutes with \(\mathbf P_M,\mathbf J_M,\mathbf H_M\). Hence the complete magnetic Killing algebra decomposes as
\begin{equation}
    \mathfrak{g}_{\rm mag}
    =
    \mathfrak{sl}(2,\mathbb R)\oplus\mathbb R.
\end{equation}
Thus, the general Killing vector \eqref{eq:General-Killing-Magnetic-Sector} is an arbitrary linear combination of a Harrison-sector element and the additional sectorial scaling direction. 

The finite magnetic Harrison map is constructed by integrating the flow generated by \(\mathbf H_M\). Using an affine parameter \(\lambda\), the evolution equations become \(\dot f=f\chi\), \(\dot\chi=f\kappa^2+\gamma\chi^2\), and \(\dot\kappa=\beta\kappa\chi\). Introducing \(R=f\kappa^2\) leads to \(\dot R=(1+2\beta)R\chi=4\gamma R\chi\), with \(2\gamma=\beta+1/2\). Defining \(X^2=R/\gamma\), the combinations \(u=\chi+X\) and \(v=\chi-X\) obey \(\dot u=\gamma u^2\) and \(\dot v=\gamma v^2\). Hence, their solutions are \(u(\lambda)=u/(1-\gamma\lambda u)\) and \(v(\lambda)=v/(1-\gamma\lambda v)\). Re-expressing the result in terms of \((f,\chi,\kappa)\), and introducing \(\mathtt p=2\gamma\lambda\), we obtain
\begin{equation}
    \Delta_M
    =
    1-\mathtt p\chi+
    \frac{\mathtt p^2}{4}
    \left(
        \chi^2-\frac{f\kappa^2}{\gamma}
    \right).
\end{equation}
Therefore the finite magnetic Harrison map is
\begin{equation}\label{eq:Mapeos-HarrisonMagnetico-M}
    \begin{aligned}
        &f'
        =
        f\Delta_M^{-1/(2\gamma)},
        \qquad
        \kappa'
        =
        \kappa\Delta_M^{-\beta/(2\gamma)},
        \\
        &\chi'
        =
        \frac{
            \chi-\frac{\mathtt p}{2}
            \left(
                \chi^2-\frac{f\kappa^2}{\gamma}
            \right)
        }{\Delta_M},
        \qquad
        \epsilon'=0,
        \qquad
        \psi'=0.
    \end{aligned}
\end{equation}
\subsection{Electric Harrison transformation}
The electric subspace is characterized by \(\epsilon=0,\quad \chi=0\). Then the metric is
\[
    \dd s_{\rm el}^{2}
    =
    \frac{\dd f^2}{2f^2}
    +
    \frac{2\epsz}{\alp^2\kappa^2}\dd\kappa^2
    -
    \frac{\kappa^2}{2f}\dd\psi^2.
\]
By applying the same procedure described in Sec.~\ref{SubSec:Magnetic Harrison transformation}, namely solving
\eqref{eq:Ecuacion De Killing} for \(K=A(f,\psi,\kappa)\partial_f+B(f,\psi,\kappa)\partial_\psi+C(f,\psi,\kappa)\partial_\kappa\), one obtains the general electric sector Killing vector
{\footnotesize
\begin{equation}\label{eq:General-Killing-Magnetic-Sector}
\begin{aligned}
K_E
={}&
f(a\psi+b)\partial_f
+
\left[
a\left(
\frac{f}{\kappa^2}+\gamma\psi^2
\right)
+
\left(
\frac{b}{2}-c
\right)\psi
+d
\right]\partial_\psi
\\
&+
\kappa(-\beta a\psi+c)\partial_\kappa .
\end{aligned}
\end{equation}
}

Here \(a,b,c,d\) are integration constants, and using the definition \eqref{eq:beta-gamma-magnetic}. Consequently, the complete electric Killing algebra is four-dimensional. The Harrison subalgebra is obtained by selecting the three directions generated by the shift, the nonlinear charging generator, and their commutator. Concretely, setting \(d=1\) yields \(\mathbf P_E\), setting \(a=1\) yields \(\mathbf H_E\), and \([\mathbf P_E,\mathbf H_E]\) defines \(\mathbf J_E\). Hence
\begin{subequations}\label{eq:Generadores-HarrisonElectrico-M}
    \begin{equation}
        \mathbf P_E=\partial_\psi,
    \end{equation}
    \begin{equation}
        \mathbf J_E=f\partial_f+2\gamma\psi\partial_\psi-\beta\kappa\partial_\kappa,
    \end{equation}
    \begin{equation}
        \mathbf H_E=f\psi\partial_f+
        \left(\frac{f}{\kappa^2}+\gamma\psi^2\right)\partial_\psi
        -\beta\kappa\psi\partial_\kappa.
    \end{equation}
\end{subequations}
The fourth independent Killing direction is given by the sectorial scaling
\(
    \mathbf S_E=-\psi\partial_\psi+\kappa\partial_\kappa,
\)
which is indeed a true Killing vector of the electric metric, but it does not correspond to the Harrison generator. The Harrison generator is instead \(\mathbf H_E\), because it generates a non-trivial electric potential starting from a configuration with \(\psi=0\), as \(\left.\mathbf H_E^\psi\right|_{\psi=0}=f/\kappa^2\).

The commutators of the electric Harrison sector are
\begin{equation}\label{eq:Conmutadores-HarrisonElectrico-M}
    \begin{aligned}
        &[\mathbf J_E,\mathbf P_E]=-2\gamma \mathbf P_E,
        \qquad
        [\mathbf J_E,\mathbf H_E]=2\gamma \mathbf H_E,
        \\
        &[\mathbf P_E,\mathbf H_E]=\mathbf J_E.
    \end{aligned}
\end{equation}
Therefore, for \(\gamma\neq0\), one has
\(
    \langle \mathbf P_E,\mathbf J_E,\mathbf H_E\rangle
    \cong
    \mathfrak{sl}(2,\R),
\)
after an appropriate rescaling. As in the magnetic case, the complete electric-sector Killing algebra is
\[
    \mathfrak g_{\rm el}
    =
    \mathfrak{sl}(2,\R)\oplus\mathbb R,
\]
where the extra \(\mathbb R\) direction is generated by the corresponding central combination of \(\mathbf S_E\) and \(\mathbf J_E\).

The finite electric Harrison map is obtained by integrating the flow of \(\mathbf H_E\). With parameter \(\mathtt q\), it is convenient to write
\begin{equation}
    \Delta_E
    =
    1-\mathtt q\psi+
    \frac{\mathtt q^2}{4}\left(\psi^2-\frac{f}{\gamma\kappa^2}\right).
\end{equation}
Therefore, the electric Harrison transformation is
\begin{equation}\label{eq:Mapeos-HarrisonElectrico-M}
    \begin{aligned}
        &f'=f\Delta_E^{-1/(2\gamma)},
        \qquad
        \kappa'=\kappa\Delta_E^{\beta/(2\gamma)},
        \\
        & \psi'
        =
        \frac{\psi-\frac{\mathtt q}{2}\left(\psi^2-\frac{f}{\gamma\kappa^2}\right)}{\Delta_E},
        \qquad
        \epsilon'=0,
        \qquad
        \chi'=0.
    \end{aligned}
\end{equation}
where \(\mathtt q \in \mathbb{R}\). This choice of parametrization is fixed so that \(\mathtt q\) generates the flow of \(\mathbf H_E/(2\gamma)\), equivalently, if \(\lambda\) denotes the affine flow parameter associated with \(\mathbf H_E\), then \(\mathtt q = 2\gamma \lambda\).

Since \(\gamma = (\alp^2 + \epsz)/(4\epsz)\), the dilatonic case with \(\epsz = +1\) exhibits no real degeneracy. In contrast, for the phantom case \(\epsz = -1\), the exceptional value corresponds to Superstring theory in the low-energy limit, where \(\alp^2 = 1\). When \(\gamma=0\), the maps \eqref{eq:Mapeos-HarrisonMagnetico-M} and \eqref{eq:Mapeos-HarrisonElectrico-M} are not valid because they contain \(1/\gamma\). However, the commutators \eqref{eq:Conmutadores-HarrisonElectrico-M} and \eqref{eq:Conmutadores-HarrisonMagnetico-M} have a well-defined contraction:
\[
    [\mathbf J,\mathbf P]=0,
    \qquad
    [\mathbf J,\mathbf H]=0,
    \qquad
    [\mathbf P,\mathbf H]=\mathbf J.
\]
Thus, in the degenerate case \(\gamma=0\) or in the \(\mathfrak{sl}(2,\R)\) sector, the Harrison contracts to the Heisenberg algebra \(\mathfrak h_3\), with \(\mathbf J\) playing the role of the central generator.

\subsection{Hidden symmetries in EMMSF}
\label{SubSec:ocultas-ModMax}
For the EMMSF theory, we observed in subsection \ref{SubSec:Einstein-ModMax-Scalar Field} that the primary modification relative to EMSF arises in the electromagnetic sector. Therefore
\paragraph{Gravitational-twist subspace (\(\psi=0,\quad \chi=0 \)):} Remains unchanged, as it does not contain any electromagnetic potentials. Consequently, \eqref{eq:Generadores-Ehlers-M}, \eqref{eq:Conmutadores-Ehlers-M}, and \eqref{eq:Mapeos-Ehlers-M} are not modified.
\paragraph{Magnetic subspace (\(\psi=0,\quad \epsilon=0 \)):} In the magnetic subspace, the metric takes the form
\[
    \dd s_{\rm mag,MM}^{2}
    =
    \frac{\dd f^2}{2f^2}
    +
    \frac{2\epsz}{\alp^2\kappa^2}\dd\kappa^2
    -
    \frac{\dd\chi^2}{2fw\kappa^2},
\]
hence, the magnetic Harrison transformation is uniquely modified by the parameter \(w\). When \(  \dd w=0 \), then  \(w=w_0\) is taken to be a constant, the resulting deformed generators are
\begin{subequations}
    \begin{align}
        \mathbf P_M^{\rm MM}&=\partial_\chi,
        \\
        \mathbf J_M^{\rm MM}
            &=
            f\partial_f
            +
            2\gamma\chi\partial_\chi
            +
            \beta\kappa\partial_\kappa,
        \\
        \mathbf H_M^{\rm MM}
            &=
            f\chi\partial_f
            +
            \left(
                w_0 f\kappa^2+\gamma\chi^2
            \right)\partial_\chi
            +
            \beta\kappa\chi\partial_\kappa.
    \end{align}
\end{subequations}
The commutation relations remain unchanged, namely \([\mathbf J_M^{\rm MM},\mathbf P_M^{\rm MM}]= -2\gamma \mathbf P_M^{\rm MM} \), \([\mathbf J_M^{\rm MM},\mathbf H_M^{\rm MM}] =  2\gamma \mathbf H_M^{\rm MM} \), and \([\mathbf P_M^{\rm MM},\mathbf H_M^{\rm MM}] =  \mathbf J_M^{\rm MM}\). Consequently, when \(\gamma\neq0\), the algebra is again isomorphic to \(\mathfrak{sl}(2,\R)\), whereas for \(\gamma=0\) it is isomorphic to \(\mathfrak{h}_3\). \emph{The maps \eqref{eq:Mapeos-HarrisonMagnetico-M} are modified in the following manner, \( f\kappa^2 \quad\longmapsto\quad w_0 f\kappa^2 \)}, yielding
{\footnotesize
\begin{equation}\label{eq:Mapeos-HarrisonMagnetico-MM}
    \begin{aligned}
        &\Delta_M^{\rm MM}
        =
        1-\mathtt p\chi
        +
        \frac{\mathtt p^2}{4}
        \left(
            \chi^2-\frac{w_0 f\kappa^2}{\gamma}
        \right),
        \\
        &f'
        =
        f\left(\Delta_M^{\rm MM}\right)^{-1/(2\gamma)},
        \qquad
        \kappa'
        =
        \kappa\left(\Delta_M^{\rm MM}\right)^{-\beta/(2\gamma)},
        \\
        &\chi'
        =
        \frac{
            \chi-\frac{\mathtt p}{2}
            \left(
                \chi^2-\frac{w_0 f\kappa^2}{\gamma}
            \right)
        }{
            \Delta_M^{\rm MM}
        },
        \qquad
        \epsilon'=0,
        \qquad
        \psi'=0.
    \end{aligned}
\end{equation}
}
\paragraph{Electric subspace (\(\chi=0,\quad \epsilon=0 \)):} In the electric case, the corresponding deformation contribution to the metric is expressed as \( \left( w + \frac{v^2}{w} \right)\kappa^2 \dd\psi^2 \), and thus the deformation term is 
\[
    \Lambda_E
    :=
    \frac{v^2+w^2}{w}.
\]
If \(\dd\Lambda_E=0\), then the generatoris are 
\begin{subequations}
    \begin{align}
        &\mathbf P_E^{\rm MM}=\partial_\psi,
        \\
        &\mathbf J_E^{\rm MM}
            =
            f\partial_f
            +
            2\gamma\psi\partial_\psi
            -
            \beta\kappa\partial_\kappa,
        \\
        &\mathbf H_E^{\rm MM}
            =
            f\psi\partial_f
            +
            \left(
                \frac{f}{\Lambda_{E0}\kappa^2}
                +
                \gamma\psi^2
            \right)\partial_\psi
            -
            \beta\kappa\psi\partial_\kappa.
    \end{align}
\end{subequations}
The commutators maintain their form \([\mathbf J_E^{\rm MM},\mathbf P_E^{\rm MM}] = -2\gamma \mathbf  P_E^{\rm MM} \), \( [\mathbf J_E^{\rm MM},\mathbf H_E^{\rm MM}]\),  \( [\mathbf P_E^{\rm MM},\mathbf H_E^{\rm MM}] = \mathbf J_E^{\rm MM}\), when \(\gamma\neq0\), the algebra is again isomorphic to \(\mathfrak{sl}(2,\R)\), whereas for \(\gamma=0\) it is isomorphic to \(\mathfrak{h}_3\). \emph{The maps \eqref{eq:Mapeos-HarrisonElectrico-M} are modified in the following manner, \(  \frac{f}{\kappa^2} \quad\longmapsto\quad\frac{f}{\Lambda_{E}\kappa^2} \)}, yielding
\begin{equation}\label{eq:Mapeos-HarrisonElectrico-MM}
    \begin{aligned}
        &\Delta_E^{\rm MM}
        =
        1-\mathtt q\psi+
        \frac{\mathtt q^2}{4}
        \left(
            \psi^2-\frac{f}{\gamma\Lambda_E\kappa^2}
        \right),
        \\
        &f'
        =
        f\left(\Delta_E^{\rm MM}\right)^{-1/(2\gamma)},
        \qquad
        \kappa'
        =
        \kappa\left(\Delta_E^{\rm MM}\right)^{\beta/(2\gamma)},
        \\
        &\psi'
        =
        \frac{
            \psi-\frac{\mathtt q}{2}
            \left(
                \psi^2-\frac{f}{\gamma\Lambda_{E}\kappa^2}
            \right)
        }{
            \Delta_E^{\rm MM}
        },
        \qquad
        \epsilon'=0,
        \qquad
        \chi'=0.
    \end{aligned}
\end{equation}
\paragraph{Simultaneous coexistence of sectorial Harrison transformations:} In order for both of Harrison's transformations to exist, the following conditions must be fulfilled
\[
    \dd w=0, \qquad \dd\left(   \frac{v^2+w^2}{w} \right)=0.
\]
Therefore, since \(\dd w = 0\), the second condition implies \(\dd(v^2) = 0\), which in turn gives \(\dd v = 0\) for \(v \neq 0\). This corresponds to the \emph{frozen ModMax sector}.

\paragraph{Relation with the Bokulić--Herdeiro Harrison maps.}
The sectorial Harrison transformations employed in this work coincide with the generalized Harrison transformations introduced by Bokulić and Herdeiro \cite{Bokulic:2025ucy} for the Einstein–ModMax and Einstein–ModMax–Dilaton frameworks, considering a dilatonic scalar field. To prevent ambiguity, we use \(\Gamma\) to denote the ModMax deformation parameter appearing in their analysis, whereas \(\gamma\) is reserved for our target-space constant defined in \eqref{eq:beta-gamma-magnetic}. In the purely electric and purely magnetic ModMax subsectors considered by Bokulić and Herdeiro, the ModMax deformation manifests itself as a constant rescaling of the electromagnetic Ernst potential. Consequently, in the electric sector one can identify
\[
    \widehat{\Phi}_{E}
    =
    e^{\Gamma/2}\Phi_{\rm BH}
    \quad\longleftrightarrow\quad
    \psi ,
\]
whereas in the magnetic sector
\[
    \widehat{\Phi}_{M}
    =
    e^{-\Gamma/2}\Phi_{\rm BH}
    \quad\longleftrightarrow\quad
    \chi .
\]
For the dilatonic extension, the scalar dictionary is
\[
    \kappa=e^{-\alpha_0\phi_{\rm BH}},
    \qquad
    \kappa^2=e^{-2\alpha_0\phi_{\rm BH}}.
\]
Therefore, the characteristic dilatonic combinations appearing in our
sectorial Harrison denominators become
\[
    \frac{f}{\kappa^2}
    =
    f e^{2\alpha_0\phi_{\rm BH}}
    \qquad
    \text{for the electric branch},
\]
and
\[
    f\kappa^2
    =
    f e^{-2\alpha_0\phi_{\rm BH}}
    \qquad
    \text{for the magnetic branch}.
\]
In the ordinary dilaton case \(\epsilon_0=+1\), our constants reduce to
\[
    \beta=\frac{\alpha_0^2}{2},
    \qquad
    \gamma=\frac{1+\alpha_0^2}{4}.
\]
Hence the electric Harrison map gives
\[
    f' = f\Delta_E^{-1/(2\gamma)}
    =
    f\Delta_E^{-2/(1+\alpha_0^2)},
\]
and
\[
    \kappa'
    =
    \kappa\Delta_E^{\beta/(2\gamma)}
    =
    \kappa\Delta_E^{\alpha_0^2/(1+\alpha_0^2)}.
\]
Equivalently,
\[
    e^{-2\alpha_0\phi_{\rm BH}'}
    =
    e^{-2\alpha_0\phi_{\rm BH}}
    \Delta_E^{2\alpha_0^2/(1+\alpha_0^2)}.
\]
Thus, in the non-phantom scenario, the scalar and metric weights coincide with the Einstein-ModMax-Dilaton Harrison weights of Bokulić-Herdeiro. Their Harrison factor has the schematic form
\[
    \lambda_{\rm BH}
    =
    1+
    \frac14 B^2(1+\alpha_0^2)
    e^{-\Gamma}
    f e^{2\alpha_0\phi_{\rm BH}}
\]
for the electric branch with vanishing seed gauge potential, while our electric Harrison denominator is
\[
    \Delta_E
    =
    1-
    \frac{q_H^2}{4\gamma}
    \frac{f}{\kappa^2}
    =
    1-
    \frac{q_H^2}{4\gamma}
    f e^{2\alpha_0\phi_{\rm BH}}.
\]
These are the same dilatonic Harrison factor up to the conventional sign or analytic continuation of the Harrison parameter, with the parameter matching
\[
    -\,\frac{q_H^2}{4\gamma}
    \quad\longleftrightarrow\quad
    \frac14 B^2(1+\alpha_0^2)e^{-\Gamma}.
\]
The magnetic branch is obtained by the dual replacement
\[
    \frac{f}{\kappa^2}
    \longrightarrow
    f\kappa^2,
    \qquad
    e^{-\Gamma}
    \longleftrightarrow
    e^{+\Gamma},
\]
or, equivalently,
\[
    f e^{2\alpha_0\phi_{\rm BH}}
    \longrightarrow
    f e^{-2\alpha_0\phi_{\rm BH}}.
\]
Accordingly, the Bokulić–Herdeiro Harrison transformations can be obtained from our sectorial Harrison maps by specializing to the purely electric or purely magnetic, dilaton sector,
\[
    \epsilon_0=+1.
\]

\subsection{Global hidden symmetries and the weight obstruction}
\label{subsec:global-hidden-obstruction}

We now explain why, for generic scalar couplings, the Ehlers and Harrison maps obtained above are only sectorial, and identify the exceptional conditions under which they become global hidden symmetries of the full target space. In stationary Einstein–Maxwell–dilaton sigma models, this is standard: after dimensional reduction, the target-space Killing algebra is generically merely solvable, and larger hidden symmetry groups appear only at critical couplings \cite{Kinnersley:1977pg,Kinnersley:1977ph,Kinnersley:1978pz,Galtsov:1995mb}.

The point can be seen directly from the target metric. Introduce the logarithmic variables
\(
    \mathtt u=\ln f,
    \quad
    \mathtt v=\ln\kappa .
\)
The Cartan, or diagonal, part of the EMSF target metric is
\(
    d\Sigma_{\rm Cartan}^2
    =
    \frac14\,\dd \mathtt u^2
    +
    \frac{\epsilon_0}{\alpha_0^2}\,\dd \mathtt v^2 ,
\)
thus the inverse Cartan metric is
\(
    G_{\rm Cartan}^{-1}
    =
    \operatorname{diag}
    \left(
        4,
        \frac{\alpha_0^2}{\epsilon_0}
    \right),
\)
this inverse metric is what defines the scalar product between the weights below.

The remaining potentials \(\epsilon,\psi,\chi\) appear in the target metric with exponential factors
\[
    e^{-2 \mathtt u},\qquad
    e^{-\mathtt u+2\mathtt v},\qquad
    e^{-\mathtt u-2\mathtt v}.
\]
We thus assign them the weight vectors
\[
    \zeta_\epsilon=(-2,0),\qquad
    \zeta_\psi=(-1,2),\qquad
    \zeta_\chi=(-1,-2),
\]
which are just the coefficients of \(\mathtt u\) and \(\mathtt v\) in the exponentials multiplying the axionic directions, not additional assumptions. For two weights \(\zeta=(\zeta_1,\zeta_2)\) and \(\eta=(\eta_1,\eta_2)\), their scalar product, defined via the inverse Cartan metric, is
\[
    \langle \zeta,\eta\rangle
    =
    4\zeta_1 \eta_1+\frac{\alpha_0^2}{\epsilon_0}\zeta_2 \eta_2.
\]

The 1-form \(\Theta \) shows that electric and magnetic translations do not close separately: their commutator yields the twist translation. At the level of weights,
\[
    \zeta_\psi+\zeta_\chi=\zeta_\epsilon .
\]
Therefore, if these three axionic directions are to be embedded into a non-Abelian global semisimple hidden symmetry, the natural minimal possibility is that they form an \(A_2\) root pattern, namely the root system associated with \(\mathfrak{sl}(3,\mathbb R)\). In that case the corresponding Cartan integer
\[
    2
    \frac{
        \langle \zeta_\psi,\zeta_\chi\rangle
    }{
        \langle \zeta_\psi,\zeta_\psi\rangle
    }
\]
must be equal to \(-1\). This is the usual root-space condition: Cartan integers are integral because they count the finite root strings of the Lie algebra \cite{Knapp:1996}.

We now compute the condition explicitly. First, imposing the \(A_2\) value gives
\[
    2
    \frac{
        \langle \zeta_\psi,\zeta_\chi\rangle
    }{
        \langle \zeta_\psi,\zeta_\psi\rangle
    }
    =
    2
    \frac{
        1-\frac{\alpha_0^2}{\epsilon_0}
    }{
        1+\frac{\alpha_0^2}{\epsilon_0}
    } =
    -1,
\] thus \(\alpha_0^2 = 3\epsilon_0\), therefore, this equality holds for real values of \(\alpha_0\) only in Kaluza–Klein theory \(\alpha_0^2=3\) when a dilatonic scalar field \(\epsilon_0=+1\) is considered. At this point the weights form an \(A_2\) root system, and the global hidden symmetry enlarges to the Kaluza–Klein \(SL(3,\mathbb R)\) symmetry \cite{Matos:1986, MATOS1988423}.

There is another special limiting case, \( \alpha_0 = 0 \). One has
\(
    \frac{\dd\kappa}{\kappa}
    =
    -\alpha_0\,\dd\phi 
\), therefore \(\dd\kappa=0\) whenever either \(\alpha_0=0\) or \(\dd\phi=0\). In both cases \(\kappa\) no longer provides a dynamical electromagnetic weight. If \(\dd\phi=0\), the theory reduces on the constant-scalar subspace to the standard Einstein-Maxwell target, and one recovers the usual hidden symmetry \(SU(2,1)\). If instead \(\alpha_0=0\) while the scalar is kept as a free spectator, the target space splits into the Einstein-Maxwell target times a free scalar direction. In that case \(SU(2,1)\) acts on the Einstein-Maxwell factor, with an additional spectator-scalar symmetry \cite{Galtsov:1995mb}.


In the frozen sector the ModMax functions are constants, \(v=v_0, w=w_0, \) and the electric and magnetic Harrison blocks are deformed as
\(
    \frac{f}{\kappa^2}
    \longmapsto
    \frac{f}{\Lambda_E\kappa^2},
    \quad
    f\kappa^2
    \longmapsto
    w_0f\kappa^2,
\) equivalently, the electromagnetic part of the target metric has the constant deformation
{\footnotesize
\[
    -\frac{1}{4f}
    \left[
        \kappa^2 \dd\psi^2
        +
        \frac{1}{\kappa^2}\dd\chi^2
    \right]
    \quad\longmapsto\quad
    -\frac{1}{4f}
    \left[
        \Lambda_E\kappa^2 \dd\psi^2
        +
        \frac{1}{w_0\kappa^2}\dd\chi^2
    \right].
\]
}
Since \(w_0\) and \(\Lambda_E\) are constants, they do not change the exponential weights in the Cartan variables \(\mathtt u\) and \(\mathtt v \). Thus the weight vectors
\(
    \zeta_\psi=(-1,2),
    \quad
    \zeta_\chi=(-1,-2)
\)
are the same as in the EMSF case. 
The possible \(A_2\) enhancement therefore still requires the same critical scalar coupling,  \(\epsilon_0=+1,\quad \alpha_0^2=3\). 

However, preserving the weights is not sufficient. A global hidden symmetry also requires that the electromagnetic block be reducible to the standard Maxwell normalization by a constant change of electromagnetic potentials compatible with the structure \(\Theta=\dd \epsilon-\psi\,\dd\chi\). Let
\( \psi=\frac{1}{\mathbf{a}}\Psi,
    \quad
    \chi=\frac{1}{\mathbf b}X 
\), then the electromagnetic block becomes
\(
    -\frac{1}{4f}
    \left[
        \frac{\Lambda_E}{\mathbf a^2}\kappa^2 \dd\Psi^2
        +
        \frac{1}{w_0\mathbf b^2\kappa^2}\dd X^2
    \right]
\), to recover the standard normalization one needs 
\(
    \mathbf a^2=\Lambda_E,
    \quad
    \mathbf b^2=\frac{1}{w_0}.
\)

But the same redefinition must preserve the 1-form \(\Theta\), then
\(
    \psi\,d\chi
    =
    \frac{1}{\mathbf a \mathbf b}\,\Psi\,dX 
\), therefore the structure keeps its canonical form only if
\[
    \mathbf a \mathbf b=1, \quad \text{implies}\quad \mathbf a^2\mathbf b^2=\frac{\Lambda_E}{w_0}=\frac{v_0^2+w_0^2}{w_0^2}=1,
\] this condition implies \(v_0=0\).

Therefore, while the frozen ModMax deformation leaves the Cartan weights unchanged, any nonzero \(v_0\) modifies the relative normalization between the electric and magnetic sectors in a way that conflicts with the underlying structure, thereby breaking the hidden global symmetry. Hence, to preserve the hidden global symmetry in the Kaluza–Klein setup with a dilaton, one must impose \(v_0=0\), and \(w_0\neq 0\). In equivalent terms, insisting on the global hidden symmetry restores the frozen ModMax sector to the Maxwell-normalizable form of linear electrodynamics.
case.

\section{Lagrangian, Hamiltonian and Noether carges}
\label{sec:hamiltoniano-noether-target}
\subsection{Lagrangian}
Considering the trajectory in the potential space
\[
\iota\longmapsto
\left(
f(\iota),\epsilon(\iota),\psi(\iota),\chi(\iota),\kappa(\iota)
\right),
\]
where the dot denotes the derivative with respect to the parameter \(\iota\). Taking into account \eqref{eq:metric-EMS}, the geodesic Lagrangian for the target is
\[
\mathcal L_{\rm M}
=
\dfrac{1}{2}
G_{AB}\dot Y^A\dot Y^B,
\qquad
Y^A=(f,\epsilon,\psi,\chi,\kappa).
\]
Or explicitly,
{\footnotesize
\begin{equation}\label{eq:Lagrangiano-M}
    \mathcal L_{\rm M}
    =
    \dfrac{1}{4f^2}
    \left[
    \dot f^2+
    \left(\dot\epsilon-\psi\dot\chi\right)^2
    \right]
    +
    \dfrac{\epsilon_0}{\alpha_0^2\kappa^2}\dot\kappa^2
    -
    \dfrac{\kappa^2}{4f}\dot\psi^2
    -
    \dfrac{1}{4f\kappa^2}\dot\chi^2.
\end{equation}
}

And the canonical momenta \(p_A=\partial L/\partial \dot Y^A\) are
\begin{subequations}\label{eq:Momenta-M}
    \begin{align}
        &p_f=\dfrac{\partial \mathcal L_{\rm M}}{\partial \dot f}
        =
        \dfrac{\dot f}{2f^2}, 
        \qquad 
        p_\epsilon=\dfrac{\partial \mathcal L_{\rm M}}{\partial \dot\epsilon}
        =
        \dfrac{\dot\epsilon-\psi\dot\chi}{2f^2},
        \\
        &p_\psi=\dfrac{\partial \mathcal L_{\rm M}}{\partial \dot\psi}
        =
        -\dfrac{\kappa^2}{2f}\dot\psi,
        \qquad 
        p_\kappa=\dfrac{\partial \mathcal L_{\rm M}}{\partial \dot\kappa}
        =
        \dfrac{2\epsilon_0}{\alpha_0^2\kappa^2}\dot\kappa
        \\
        &p_\chi=\dfrac{\partial \mathcal L_{\rm M}}{\partial \dot\chi}
        =
        -\dfrac{\psi}{2f^2}
        \left(
        \dot\epsilon-\psi\dot\chi
        \right)
        -
        \dfrac{1}{2f\kappa^2}\dot\chi.
    \end{align}
\end{subequations}

\paragraph{ModMax case:} For the ModMax scenario, the lagrangian is
{\footnotesize
\begin{equation}\label{eq:Lagrangiano-MM}
\begin{aligned}
    \mathcal L_{\rm MM}
    =
    \frac{1}{4f^2}\left[\dot f^2+(\dot\epsilon-\psi\dot\chi)^2\right]
    +
    \frac{\epsz}{\alp^2\kappa^2}\dot\kappa^2
    \\
    -
    \frac{1}{4f}\left[
    \frac{v^2+w^2}{w}\kappa^2\dot\psi^2
    +
    \frac{1}{w\kappa^2}\dot\chi^2
    +2\frac{v}{w}\dot\psi\dot\chi
    \right].
\end{aligned}
\end{equation}
}
\subsection{Hamiltonian}
Solving for the potentials in terms of the moments, we obtain
\begin{subequations}\label{eq:PotencialesEnTerminosDeMomentos}
    \begin{align}
    &\dot f=2f^2p_f,
    \qquad
    \dot\epsilon-\psi\dot\chi=2f^2p_\epsilon,
    \\
    &\dot\psi=-\frac{2f}{\kappa^2}p_\psi,
    \qquad
    \dot\chi=-2f\kappa^2(p_\chi+\psi p_\epsilon),
    \\
    &\dot\kappa=\frac{\alp^2\kappa^2}{2\epsz}p_\kappa.
    \end{align}
\end{subequations}
Therefore, the geodesic Hamiltonian \(\mathcal H=\frac12G^{AB}p_Ap_B\) takes the form
{\footnotesize
\begin{equation}\label{eq:H-EMS}
    \mathcal H_{\rm M}
    =
     f^2p_f^2
    +f^2p_\epsilon^2
    -\frac{f}{\kappa^2}p_\psi^2
    -f\kappa^2(p_\chi+\psi p_\epsilon)^2
    +\frac{\alp^2\kappa^2}{4\epsz}p_\kappa^2.
\end{equation}
}
\paragraph{ModMax case}
Starting from the Lagrangian \eqref{eq:Lagrangiano-MM}, only the electromagnetic moments change to 
\begin{equation}\label{eq:MomentosElectromagneticos-MM}
\begin{aligned}
    p_\psi
    =-\frac{1}{2f}\left[
    \frac{v^2+w^2}{w}\kappa^2\dot\psi
    +\frac{v}{w}\dot\chi
    \right],
    \\
    p_\chi
    =-\psi p_\epsilon
    -\frac{1}{2f}\left[
    \frac{v}{w}\dot\psi
    +\frac{1}{w\kappa^2}\dot\chi
    \right].
\end{aligned}
\end{equation}
We set \(\Pi_\chi := p_\chi + \psi\, p_\epsilon\), and the determinant of the electromagnetic momentum matrix \eqref{eq:MomentosElectromagneticos-MM} is given by
\(
\left(\frac{v^2 + w^2}{w}\kappa^2\right)\left(\frac{1}{w\kappa^2}\right) - \left(\frac{v}{w}\right)^2 = 1,
\)
and thus
\begin{equation}
    \begin{aligned}
        \dot\psi=-2f\left[\frac{1}{w\kappa^2}p_\psi-\frac{v}{w}\Pi_\chi\right],
        \\
        \dot\chi=-2f\left[-\frac{v}{w}p_\psi+\frac{v^2+w^2}{w}\kappa^2\Pi_\chi\right].
    \end{aligned}
\end{equation}
The general ModMax geodesic Hamiltonian is
\begin{equation}\label{eq:H-MM-general}
\begin{aligned}
\mathcal H_{\rm MM}
=&
 f^2p_f^2
+f^2p_\epsilon^2
+\frac{\alp^2\kappa^2}{4\epsz}p_\kappa^2
\\
&-f\left[
\frac{1}{w\kappa^2}p_\psi^2
-2\frac{v}{w}p_\psi\Pi_\chi
+\frac{v^2+w^2}{w}\kappa^2\Pi_\chi^2
\right].
\end{aligned}
\end{equation}

\subsection{Noether charges and sectorial Casimirs}
If \(X=X^A\partial_A\) is a Killing vector of the relevant target metric, then \(\Q_X=X^A p_A\) is conserved. Indeed, using \(\nabla_{(A}X_{B)}=0\), one obtains
\[
    \frac{\dd \Q_X}{\dd\lambda}
    =
    \frac{\dd}{\dd\lambda}(X_A\dot Y^A)
    =
    \nabla_{(A}X_{B)}\dot Y^A\dot Y^B=0,
\]
or equivalently \(\{\Q_X,H\}=0\). Moreover, the Poisson brackets reproduce the Killing algebra, \(\{\Q_X,\Q_Y\}=\Q_{[X,Y]}\). 

At the algebraic level, the Casimir is the quadratic element built from a basis of the Lie algebra which commutes with all generators. For the Ehlers basis \((\mathbf T_E,\mathbf D_E,\mathbf E_E)\), satisfying
\(
    [\mathbf D_E,\mathbf T_E]=-\mathbf T_E,
    \quad
    [\mathbf D_E,\mathbf E_E]=\mathbf E_E,
    \quad
    [\mathbf T_E,\mathbf E_E]=2\mathbf D_E,
\)
the quadratic classical Casimir \(\mathbf D_E^2-\frac12(\mathbf T_E\mathbf E_E+\mathbf E_E\mathbf T_E)\) is
\[
    \C_{\rm grav}=\Q_D^2-\Q_T\Q_E.
\]
For a Harrison basis \((\mathbf P,\mathbf J,\mathbf H)\), satisfying
\(
    [\mathbf J,\mathbf P]=-2\gamma\mathbf P,
    \quad
    [\mathbf J,\mathbf H]=2\gamma\mathbf H,
    \quad
    [\mathbf P,\mathbf H]=\mathbf J,
\)
the quadratic classical Casimir \(\mathbf J^2-2\gamma(\mathbf P\mathbf H+\mathbf H\mathbf P)\) is
\[
    \C=\Q_J^2-4\gamma\Q_P\Q_H.
\]

\subsubsection{Visible charges}
\begin{equation}\label{eq:CargasNoether-Visibles}
    \begin{aligned}
    \Q_{K_\epsilon}
    &=p_\epsilon
    =\frac{\dot\epsilon-\psi\dot\chi}{2f^2},
    \\
    \Q_{K_\chi}
    &=p_\chi
    =
    -\frac{\psi(\dot\epsilon-\psi\dot\chi)}{2f^2}
    -\frac{\dot\chi}{2f\kappa^2},
    \\
    \Q_{K_\psi}
    &=p_\psi+\chi p_\epsilon
    =
    -\frac{\kappa^2\dot\psi}{2f}
    +
    \frac{\chi(\dot\epsilon-\psi\dot\chi)}{2f^2},
    \\
    \Q_{D_g}
    &=
    fp_f+\epsilon p_\epsilon+\frac12\psi p_\psi+\frac12\chi p_\chi
    \\
    &=
    \frac{\dot f}{2f}
    +
    \frac{\epsilon(\dot\epsilon-\psi\dot\chi)}{2f^2}
    -\frac{\psi\kappa^2\dot\psi}{4f}
    -\frac{\chi\psi(\dot\epsilon-\psi\dot\chi)}{4f^2}
    -\frac{\chi\dot\chi}{4f\kappa^2},
    \\
    \Q_{D_s}
    &=
    \kappa p_\kappa+\chi p_\chi-\psi p_\psi
    \\
    &=
    \frac{\dot\kappa}{\beta\kappa}
    -\frac{\chi\psi(\dot\epsilon-\psi\dot\chi)}{2f^2}
    -\frac{\chi\dot\chi}{2f\kappa^2}
    +\frac{\psi\kappa^2\dot\psi}{2f}.
    \end{aligned}
\end{equation}

\subsubsection{Ehlers gravitational scenario}

The Ehlers charges are
\begin{equation}\label{eq:CargasNoether-Ehlers}
    \begin{aligned}
    \Q_T
    &=p_\epsilon
    =\frac{\dot\epsilon}{2f^2},
    \\
    \Q_D
    &=fp_f+\epsilon p_\epsilon
    =
    \frac{\dot f}{2f}
    +
    \frac{\epsilon\dot\epsilon}{2f^2},
    \\
    \Q_E
    &=
    2\epsilon f p_f+(\epsilon^2-f^2)p_\epsilon
    =
    \frac{\epsilon\dot f}{f}
    +
    \frac{(\epsilon^2-f^2)\dot\epsilon}{2f^2}.
    \end{aligned}
\end{equation}
The Casimir is
\begin{equation}\label{eq:Casimir-Ehlers}
\begin{aligned}
\C_{\rm grav}
&=\Q_D^2-\Q_T\Q_E
=f^2(p_f^2+p_\epsilon^2)
=
\frac{\dot f^2+\dot\epsilon^2}{4f^2}
.
\end{aligned}
\end{equation}

\subsubsection{Magnetic Harrison scenario}
The magnetic Harrison charges are
\begin{equation}\label{eq:CargasNoether-Harrison-M}
    \begin{aligned}
    \Q_{P_M}
    &=p_\chi
    =
    -\frac{\dot\chi}{2f\kappa^2},
    \\
    \Q_{J_M}
    &=
    fp_f+2\gamma\chi p_\chi+\beta\kappa p_\kappa
    =
    \frac{\dot f}{2f}
    -\frac{\gamma\chi\dot\chi}{f\kappa^2}
    +\frac{\dot\kappa}{\kappa},
    \\
    \Q_{H_M}
    &=
    f\chi p_f+(f\kappa^2+\gamma\chi^2)p_\chi+\beta\kappa\chi p_\kappa
    \\
    &=
    \frac{\chi\dot f}{2f}
    -\frac{\dot\chi}{2}
    -\frac{\gamma\chi^2\dot\chi}{2f\kappa^2}
    +\frac{\chi\dot\kappa}{\kappa}.
    \end{aligned}
\end{equation}
The Casimir is
\begin{equation}\label{eq:Casimir-Harrison-M}
    \begin{aligned}
    \C_M
    &=\Q_{J_M}^2-4\gamma\Q_{P_M}\Q_{H_M}
    \\
    &=
    f^2p_f^2
    -4\gamma f\kappa^2p_\chi^2
    +\beta^2\kappa^2p_\kappa^2
    +2\beta f\kappa p_fp_\kappa
    \\
    &=
    \frac{\dot f^2}{4f^2}
    -\frac{\gamma\dot\chi^2}{f\kappa^2}
    +\frac{\dot\kappa^2}{\kappa^2}
    +\frac{\dot f\dot\kappa}{f\kappa}
    .
    \end{aligned}
\end{equation}

\subsubsection{Electric Harrison scenario}
The electric Harrison charges are
\begin{equation}\label{eq:CargasNoether-Harrison-E}
    \begin{aligned}
    \Q_{P_E}
    &=p_\psi
    =
    -\frac{\kappa^2\dot\psi}{2f},
    \\
    \Q_{J_E}
    &=
    fp_f+2\gamma\psi p_\psi-\beta\kappa p_\kappa
    =
    \frac{\dot f}{2f}
    -\frac{\gamma\kappa^2\psi\dot\psi}{f}
    -\frac{\dot\kappa}{\kappa},
    \\
    \Q_{H_E}
    &=
    f\psi p_f+
    \left(\frac{f}{\kappa^2}+\gamma\psi^2\right)p_\psi
    -\beta\kappa\psi p_\kappa
    \\
    &=
    \frac{\psi\dot f}{2f}
    -\frac{\dot\psi}{2}
    -\frac{\gamma\kappa^2\psi^2\dot\psi}{2f}
    -\frac{\psi\dot\kappa}{\kappa}.
    \end{aligned}
\end{equation}
The Casimir is
\begin{equation}\label{eq:Casimir-Harrison-E}
    \begin{aligned}
    \C_E
    &=\Q_{J_E}^2-4\gamma\Q_{P_E}\Q_{H_E}
    \\
    &=
    f^2p_f^2
    -\frac{4\gamma f}{\kappa^2}p_\psi^2
    +\beta^2\kappa^2p_\kappa^2
    -2\beta f\kappa p_fp_\kappa
    \\
    &=
    \frac{\dot f^2}{4f^2}
    -\frac{\gamma\kappa^2\dot\psi^2}{f}
    +\frac{\dot\kappa^2}{\kappa^2}
    -\frac{\dot f\dot\kappa}{f\kappa}
    .
    \end{aligned}
\end{equation}

\subsubsection{ModMax Harrison scenario}
\paragraph{Ehlers gravitational scenario:} Since the electromagnetic sector is switched off in this scenario, the formulas \eqref{eq:CargasNoether-Ehlers} and \eqref{eq:Casimir-Ehlers} remain the same.

In the frozen ModMax branch, where \(v\) and \(w\) are constants, the Harrison algebras keep the same commutation relations as in the Maxwell case. However, the electromagnetic charges must be evaluated using the momenta \eqref{eq:MomentosElectromagneticos-MM}.

\paragraph{Magnetic Harrison scenario:} In the magnetic sector, \(\psi=0\) and \(\dot\psi=0\). Hence
\(
    p_\chi
    =
    -\frac{\dot\chi}{2fw\kappa^2},
    \quad
    p_\psi
    =
    -\frac{v}{2fw}\dot\chi
    =
    v\kappa^2p_\chi.
\)
The magnetic Harrison charges are therefore
\begin{equation}\label{eq:CargasNoether-Harrison-MM-M}
\begin{aligned}
    \Q_{P_M}^{\rm MM}
    &=
    p_\chi
    =
    -\frac{\dot\chi}{2fw\kappa^2},
    \\
    \Q_{J_M}^{\rm MM}
    &=
    fp_f+2\gamma\chi p_\chi+\beta\kappa p_\kappa
    =
    \frac{\dot f}{2f}
    -
    \frac{\gamma\chi\dot\chi}{fw\kappa^2}
    +
    \frac{\dot\kappa}{\kappa},
    \\
    \Q_{H_M}^{\rm MM}
    &=
    f\chi p_f+
    (wf\kappa^2+\gamma\chi^2)p_\chi+
    \beta\kappa\chi p_\kappa
    \\
    &=
    \frac{\chi\dot f}{2f}
    -
    \frac{\dot\chi}{2}
    -
    \frac{\gamma\chi^2\dot\chi}{2fw\kappa^2}
    +
    \frac{\chi\dot\kappa}{\kappa}.
\end{aligned}
\end{equation}
The magnetic ModMax Casimir is
\begin{equation}\label{eq:Casimir-Harrison-MM-M}
\begin{aligned}
    \C_M^{\rm MM}
    &=
    \left(\Q_{J_M}^{\rm MM}\right)^2
    -
    4\gamma\Q_{P_M}^{\rm MM}\Q_{H_M}^{\rm MM}
    \\
    &=
    f^2p_f^2
    -
    4\gamma wf\kappa^2p_\chi^2
    +
    \beta^2\kappa^2p_\kappa^2
    +
    2\beta f\kappa p_fp_\kappa
    \\
    &=
    \frac{\dot f^2}{4f^2}
    -
    \frac{\gamma\dot\chi^2}{fw\kappa^2}
    +
    \frac{\dot\kappa^2}{\kappa^2}
    +
    \frac{\dot f\dot\kappa}{f\kappa}.
\end{aligned}
\end{equation}

\paragraph{Electric Harrison scenario:}In the electric sector, \(\chi=0\) and \(\dot\chi=0\). Therefore
\(
    p_\psi
    =
    -\frac{1}{2f}
    \frac{v^2+w^2}{w}\kappa^2\dot\psi,
    \quad
    \Pi_\chi
    =
    -\frac{v}{2fw}\dot\psi.
\)
Taking into account the definition of the deformation term $\Lambda_E$, one has
\(
    p_\psi=-\frac{\Lambda_E\kappa^2\dot\psi}{2f}.
\)
The electric Harrison charges become
\begin{equation}\label{eq:CargasNoether-Harrison-MM-E}
\begin{aligned}
    \Q_{P_E}^{\rm MM}
    &=
    p_\psi
    =
    -\frac{\Lambda_E\kappa^2\dot\psi}{2f},
    \\
    \Q_{J_E}^{\rm MM}
    &=
    fp_f+2\gamma\psi p_\psi-\beta\kappa p_\kappa
    =
    \frac{\dot f}{2f}
    -
    \frac{\gamma\Lambda_E\kappa^2\psi\dot\psi}{f}
    -
    \frac{\dot\kappa}{\kappa},
    \\
    \Q_{H_E}^{\rm MM}
    &=
    f\psi p_f+
    \left(
        \frac{f}{\Lambda_E\kappa^2}
        +
        \gamma\psi^2
    \right)p_\psi
    -
    \beta\kappa\psi p_\kappa
    \\
    &=
    \frac{\psi\dot f}{2f}
    -
    \frac{\dot\psi}{2}
    -
    \frac{\gamma\Lambda_E\kappa^2\psi^2\dot\psi}{2f}
    -
    \frac{\psi\dot\kappa}{\kappa}.
\end{aligned}
\end{equation}
The electric ModMax Casimir is
\begin{equation}\label{eq:Casimir-Harrison-MM-E}
\begin{aligned}
    \C_E^{\rm MM}
    &=
    \left(\Q_{J_E}^{\rm MM}\right)^2
    -
    4\gamma\Q_{P_E}^{\rm MM}\Q_{H_E}^{\rm MM}
    \\
    &=
    f^2p_f^2
    -
    \frac{4\gamma f}{\Lambda_E\kappa^2}p_\psi^2
    +
    \beta^2\kappa^2p_\kappa^2
    -
    2\beta f\kappa p_fp_\kappa
    \\
    &=
    \frac{\dot f^2}{4f^2}
    -
    \frac{\gamma\Lambda_E\kappa^2\dot\psi^2}{f}
    +
    \frac{\dot\kappa^2}{\kappa^2}
    -
    \frac{\dot f\dot\kappa}{f\kappa}.
\end{aligned}
\end{equation}

\section{Harmonic-dual projection}
\label{sec:harmonic-dual-noether-quadratures}
In this section, we set up the harmonic sector of the reduced sigma-model presented in  \cite{Bixano:2026xum,Bixano:2026ouq,Matos:2000za,Matos:2000ai,Matos:2010pcd,Matos:1995my}. The key observation is that if all target-space potentials depend on a single harmonic function,
\(
    Y^A=(f,\epsilon,\psi,\chi,\kappa)=Y^A(\mathfrak s),
    \quad
    \mathfrak s=\mathfrak s(\rho,z),
\)
then the field equations simplify to affine geodesic equations on the target space. As a result, the corresponding geodesic Hamiltonian is constant and governs the quadrature for the Weyl function \(k\). The manifest Noether charges determine the dual-harmonic quadratures for \(\omega\) and \(A_\varphi\). 

\subsection{Harmonic and dual-harmonic functions}
Following \cite{Bixano:2026xum,Bixano:2026ouq}, a harmonic function \(\mathfrak s\) satisfies
\(
    \frac{1}{\rho}D(\rho D\mathfrak s)=0,
\)
thus the associated dual harmonic function \(\widetilde{\mathfrak s}\) is then defined by
\(
    D\widetilde{\mathfrak s}
    =
    \rho\,\widetilde D\mathfrak s,
\)
then
\(
    D\mathfrak s\cdot D\widetilde{\mathfrak s}=0,
    \quad
    (D\widetilde{\mathfrak s})^2=\rho^2(D\mathfrak s)^2,
    \quad
    D\left(\frac{1}{\rho}D\widetilde{\mathfrak s}\right)=0,
\) is fulfill and where \(D=(\partial_\rho,\partial_z)^T\), \(\Tilde{D}=(\partial_z,-\partial_\rho)^T\).
Therefore, the natural basis for the harmonic branch is
\(
    \left\{
        d\mathfrak s,\,
        \frac{1}{\rho}d\widetilde{\mathfrak s}
    \right\}.
\)
The component \(d\mathfrak s\) governs the geodesic in the target space, whereas its dual \(d\widetilde{\mathfrak s}\) enters the reconstruction of the metric functions \(\omega\) and \(A_\varphi\) of the electric Weyl-Lewis-Papapetrou metric \eqref{eq:Metrica_LEP-E}, and the 4-potential \( A_{\mu} = \big[ A_t(\rho,z),\,0,\,0,\, A_{\varphi \, L}(\rho,z) \big]\).

\subsection{Projection of the 1-forms \(A,B,C\)}

Taking into account \( Y^A=Y^A(\mathfrak s)\) and denote \(d/d\mathfrak s\) by a prime. Hence \(DY^A=Y^{A\prime}D\mathfrak s\), and in particular
\(
    D\epsilon-\psi D\chi=(\epsilon'-\psi\chi')D\mathfrak s.
\)
For the 1-forms introduced in Eq. (14) of \cite{Bixano:2026xum} or, alternatively, in Eq. (19) of \cite{Bixano:2026ouq}
\begin{equation}\label{eq:1Formas-ABC}
    \begin{aligned}
        &A=
        \frac{1}{2f}
        \left[
            Df-i(D\epsilon-\psi D\chi)
        \right],
        \\
        &B=
        -\frac{1}{2\sqrt f}
        \left[
            \kappa D\psi-\frac{i}{\kappa}D\chi
        \right],
        \qquad
        C=-\frac{D\kappa}{\kappa}.
    \end{aligned}
\end{equation}

Therefore
\(
    A=a(\mathfrak s)D\mathfrak s,
    \quad
    B=b(\mathfrak s)D\mathfrak s,
    \quad
    C=c(\mathfrak s)D\mathfrak s,
\)
with
\[
    a(\mathfrak s)=
    \frac{1}{2f}
    \left[
        f'-i(\epsilon'-\psi\chi')
    \right],
\]
\[
    b(\mathfrak s)=
    -\frac{1}{2\sqrt f}
    \left[
        \kappa\psi'-\frac{i}{\kappa}\chi'
    \right],
    \qquad
    c(\mathfrak s)=-\frac{\kappa'}{\kappa}.
\]
Consequently,
\[
    a(\mathfrak s)\bar a(\mathfrak s)=
    \frac{(f')^2+(\epsilon'-\psi\chi')^2}{4f^2},
\]
\[
    b(\mathfrak s)\bar b(\mathfrak s)=
    \frac{1}{4f}
    \left[
        \kappa^2(\psi')^2+\frac{1}{\kappa^2}(\chi')^2
    \right],
    \qquad
    c(\mathfrak s)^2=\frac{(\kappa')^2}{\kappa^2}.
\]
\subsection{Reduction to a affine geodesic}
The reduced sigma-model equations, presented schematically in equation (28) of \cite{Bixano:2026xum}, take the following general form
\[
    \frac{1}{\rho}D_i(\rho D^iY^A)
    +
    \Gamma^A{}_{BC}D_iY^BD^iY^C=0.
\]
Using \(Y^A=Y^A(\mathfrak s)\), we get
\(
    D_iY^A=Y^{A\prime}D_i\mathfrak s,
\)
hence
\[
    \left(
        Y^{A\prime\prime}
        +
        \Gamma^A{}_{BC}Y^{B\prime}Y^{C\prime}
    \right)
    D_i\mathfrak sD^i\mathfrak s
    +
    Y^{A\prime}
    \frac{1}{\rho}D_i(\rho D^i\mathfrak s)=0.
\]
Since \(\mathfrak s\) is harmonic, the second term vanishes. Thus
\begin{equation}\label{eq:target-geodesic-harmonic}
    Y^{A\prime\prime}
    +
    \Gamma^A{}_{BC}Y^{B\prime}Y^{C\prime}
    =
    0.
\end{equation}
Therefore, the harmonic branch is an affinely parametrized geodesic of the target space.

\subsection{Hamiltonian constant and the quadrature of \(k\)}
If we consider a purely harmonic branch, then the Hamiltonian \eqref{eq:H-EMS}, in terms of \( a(\mathfrak s),b(\mathfrak s),c(\mathfrak s) \) takes the form

\begin{equation}\label{eq:H-EMS-abc-s}
    \mathcal H_{\rm EMS}^{(\mathfrak s)}
    =
    a(\mathfrak s)\bar a(\mathfrak s)-b(\mathfrak s)\bar b(\mathfrak s)+
    \frac{\epsilon_0}{\alpha_0^2}c(\mathfrak s)^2.
\end{equation}
Because \(Y^A(\mathfrak s)\) satisfies the geodesic equation,
{\footnotesize
\[
    \frac{d}{d\mathfrak s}
    \left(
        \frac12G_{AB}Y^{A\prime}Y^{B\prime}
    \right)
    =
    G_{AB}Y^{A\prime}
    \left(
        Y^{B\prime\prime}
        +
        \Gamma^B{}_{CD}Y^{C\prime}Y^{D\prime}
    \right)=0.
\]
}
Therefore, from \eqref{eq:target-geodesic-harmonic},  \(\mathcal H_{\rm M}(\mathfrak s) =  \mathcal K_0 = \text{constant}\). This constant is the same one that appears in the quadrature of the Weyl function \(k\) given in (79) of Appendix A in \cite{Bixano:2026xum} and in (42) of Appendix A in \cite{Bixano:2026ouq}. In the harmonic branch,
\(
    k_{,\zeta}
    =
    \mathcal K_0\,\rho\,\mathfrak s_{,\zeta}^{\,2},
\) where \(\zeta=\rho+i z\).
If \(\mathfrak R\) is defined by
\(
    \mathfrak R_{,\zeta}
    =
    \rho\,\mathfrak s_{,\zeta}^{\,2},
\)
then
\begin{equation}\label{eq:k-quadrature-harmonic}
    k=k_0+\mathcal K_0\,\mathfrak R.
\end{equation}
\emph{Thus \(\mathcal K_0\) is not an arbitrary integration constant detached from the target geometry, it is the geodesic energy of the harmonic curve.}

\subsection{Visible quadratures}

Let \(X=X^A(Y)\partial_A\) be a Killing vector of the target. Along the harmonic curve,
\(
    \mathcal Q_X
    =
    G_{AB}X^AY^{B\prime}
    =
    X^A \,p_A.
\)
Using the geodesic equation,
\(
    \frac{d\mathcal Q_X}{d\mathfrak s}
    =
    Y^{A\prime}Y^{B\prime}\nabla_{(A}X_{B)}=0,
\)
hence every target Killing vector gives a constant of the harmonic branch. From \eqref{eq:Momenta-M}, the visible Killing \(K_\epsilon=\partial_\epsilon\) gives
\(
    \mathcal Q_{K_\epsilon}
    =
    p_\epsilon
    =
    \frac{\epsilon'-\psi\chi'}{2f^2},
\)
Thus we define
\begin{equation}\label{eq:Qomega-harmonic}
    \omega_0
    :=
    2 p_\epsilon
    =
    \frac{\epsilon'-\psi\chi'}{f^2}
    =
    \text{constant}.
\end{equation}
The reconstruction equation for the rotational potential becomes form eq. (8) of \cite{Bixano:2026xum}, given by
\(
    D\omega=-\omega_0D\widetilde{\mathfrak s},
\)
so that
\begin{equation}\label{eq:omega-harmonic-quadrature}
    \omega=-\omega_0\widetilde{\mathfrak s}+\omega_1.
\end{equation}

Similarly, the visible magnetic Killing \(K_\chi=\partial_\chi\) gives
\(
    \mathcal Q_{K_\chi}=p_\chi
    =
    -\psi p_\epsilon
    -
    \frac{\chi'}{2f\kappa^2},
\)
using \(p_\epsilon=\omega_0/2\), we define the magnetic Noether charge as
\begin{equation}\label{eq:p-harmonic-EMS}
    \mathtt p:=
    -\mathcal Q_{K_\chi}
    =
    -p_\chi
    =
    \frac{\chi'}{2f\kappa^2}
    +
    \frac{\omega_0}{2}\psi
    =
    \text{constant}.
\end{equation}

The constants \(\omega_0,\mathtt p\) control the dual-harmonic quadratures for \(\omega\) and \(A_\varphi\). With the normalization used here, the angular electromagnetic potential eq. (8) of \cite{Bixano:2026xum}, is 
\begin{equation}\label{eq:Aphi-harmonic-quadrature}
    A_\varphi
    =
    A_{\varphi0}-
    \frac{\omega_1}{2}\psi
    +
     \left( \omega_0 \, \frac{\psi}{2}  - \mathtt p \right) \widetilde{\mathfrak s}
\end{equation}

\subsection{Frozen ModMax harmonic branch}

For the Frozen Einstein-ModMax-Scalar Field case the natural ModMax electromagnetic variable is eq. (19) from \cite{Bixano:2026ouq}, i.e. 
\(
    B_{\rm MM}
    =
    -\frac{1}{2\sqrt f}
    \left[
        \sqrt w_0\,\kappa D\psi
        -
        \frac{i}{\sqrt w_0\,\kappa}
        (D\chi+v_0\kappa^2D\psi)
    \right].
\)
On the harmonic branch, the projected frozen ModMax Hamiltonian is
\begin{equation}\label{eq:H-MM-abc-harmonic}
    \mathcal H_{\rm MM}^{(\mathfrak s)}
    =
    a(\mathfrak s)\bar a(\mathfrak s)-b(\mathfrak s)_{\rm MM}\bar b(\mathfrak s)_{\rm MM}
    +
    \frac{\epsilon_0}{\alpha_0^2}c(\mathfrak s)^2.
\end{equation}
If the harmonic curve solves the frozen ModMax target geodesic equations, then
\(
    \mathcal H_{\rm MM} (\mathfrak s)=\mathcal K_0^{\rm MM}
    =
    \text{constant},
\)
and this constant controls the same \(k\)-quadrature.

Thus, using the definition \eqref{eq:Qomega-harmonic}, the visible magnetic Noether charge in the frozen ModMax branch is
\begin{equation}\label{eq:pMM-harmonic-omega0}
    \mathtt p_{\rm MM}
    :=
    -\mathcal Q_{K_\chi}^{\rm MM}
    =
    -p_\chi
    =
    \frac{\chi'+v_0\kappa^2\psi'}{2fw_0\kappa^2}
    +
    \frac{\omega_0}{2}\psi
    =
    \text{constant}.
\end{equation}

Therefore, the twist quadrature \eqref{eq:omega-harmonic-quadrature} remains unchanged, whereas the \(A_\varphi\)-quadrature retains the same structural form after replacing \eqref{eq:pMM-harmonic-omega0}. \footnote{The quadratures are defined in equation (7) of Ref. \cite{Bixano:2026ouq}.}

\section{Magnetic LWP frame, Noether charges and duality map}\label{Sec:AnaBokulic}

We now consider the magnetic Lewis–Weyl–Papapetrou frame \cite{Barrientos:2023dlf}
\[
    ds^2_{\rm m}
    =
    f(d\varphi-\omega dt)^2
    +
    \frac{1}{f}
    \left[
        e^{2k}(d\rho^2+dz^2)-\rho^2dt^2
    \right].
\]
In this frame the natural electromagnetic potential is the angular one, so we write
\[
    \psi_{\rm m}=2A_\varphi^{\rm m}.
\]
The discrete map between the electric and magnetic LWP descriptions is
{\footnotesize
\[
    (f,\epsilon,\psi,\chi,\kappa)
    \longmapsto
    (f_{\rm m},\epsilon_{\rm m},\psi_{\rm m},\chi_{\rm m},\kappa_{\rm m})
    =
    \left(
        f,\,
        \epsilon-\psi\chi,\,
        \chi,\,
        -\psi,\,
        \kappa^{-1}
    \right).
\]
}
With this map one has
\(
    \dd \epsilon_{\rm m}-\psi_{\rm m}\dd  \chi_{\rm m}
    =
    \dd  \epsilon-\psi \dd \chi,
\)
and
\(
    \kappa_{\rm m}^2\dd \psi_{\rm m}^2
    +
    \kappa_{\rm m}^{-2}\dd \chi_{\rm m}^2
    =
    \kappa^{-2}\dd \chi^2+\kappa^2\dd \psi^2.
\)
Thus the LWP magnetic map is an isometry of \eqref{eq:metric-EMS}. The canonical one-form is preserved,
\(
    p_A\,\dd Y^A
    =
    p_{A_{\rm m}}\,\dd Y_{\rm m}^A,
\)
and therefore the momenta transform as
\begin{equation}
    \begin{aligned}
        &p_{f_{\rm m}}=p_f,
        \qquad
        p_{\epsilon_{\rm m}}=p_\epsilon,
        \\
        &p_{\psi_{\rm m}}=p_\chi+\psi p_\epsilon,
        \qquad
        p_{\chi_{\rm m}}=-p_\psi-\chi p_\epsilon,
        \\
        &p_{\kappa_{\rm m}}=-\kappa^2p_\kappa.
    \end{aligned}
\end{equation}
Consequently, the visible Noether charges transform as
\begin{equation}
    \begin{aligned}
        &\mathcal Q_{K_{\epsilon_{\rm m}}}
        =
        \mathcal Q_{K_\epsilon},
        \\
        &\mathcal Q_{K_{\chi_{\rm m}}}
        =
        -\mathcal Q_{K_\psi},
        \qquad
        \mathcal Q_{K_{\psi_{\rm m}}}
        =
        \mathcal Q_{K_\chi}.
    \end{aligned}
\end{equation}

Equivalently, if
\(
    \mathtt p:=-\mathcal Q_{K_\chi},
    \quad
    \mathtt q:=-\mathcal Q_{K_\psi},
\)
then
\[
    \mathtt p_{\rm m}=-\mathtt q,
    \qquad
    \mathtt q_{\rm m}=\mathtt p.
\]
Thus the magnetic LWP map exchanges electric and magnetic Noether charges, up to the conventional sign.

The twist charge is invariant: , i.e. 
\(
    \omega_{0,{\rm m}}
    :=
    2p_{\epsilon_{\rm m}}
    =
    2p_\epsilon
    =
    \omega_0,
\)
hence the rotational quadrature keeps the same form,
\(
    D\omega_{\rm m}
    =
    -\omega_0D\widetilde{\mathfrak s},
    \quad
    \omega_{\rm m}
    =
    -\omega_0\widetilde{\mathfrak s}+\omega_1.
\)

In the magnetic frame, by contrast, the potential associated directly with the target coordinate is \(A_\varphi^{\rm m}=\psi_{\rm m}/2\). The dual electromagnetic quadrature retains the same structure as in the electric frame, but is now expressed in magnetic variables:
\[
    \frac{\chi_{\rm m}'}{2f_{\rm m}\kappa_{\rm m}^2}
    =
    \mathtt p_{\rm m}
    -
    \frac{\omega_0}{2}\psi_{\rm m}.
\]
Hence, the magnetic LWP map does not introduce a new class of quadrature; rather, it simply interchanges which electromagnetic component is treated as the direct potential and which is obtained from the dual-harmonic equation.

The Ehlers generators keep their form in the magnetic variables,
\(
    \mathbf T_{\rm m}=\partial_{\epsilon_{\rm m}},
    \qquad
    \mathbf D_{\rm m}
    =
    f_{\rm m}\partial_{f_{\rm m}}
    +
    \epsilon_{\rm m}\partial_{\epsilon_{\rm m}},
\)
\(
    \mathbf E_{\rm m}
    =
    2\epsilon_{\rm m}f_{\rm m}\partial_{f_{\rm m}}
    +
    (\epsilon_{\rm m}^2-f_{\rm m}^2)\partial_{\epsilon_{\rm m}}.
\)
The Harrison sectors are exchanged. More precisely,
\[
    (\mathbf P_{E,{\rm m}},\mathbf J_{E,{\rm m}},\mathbf H_{E,{\rm m}})
    \longmapsto
    (\mathbf P_M,\mathbf J_M,\mathbf H_M),
\]
whereas
\[
    (\mathbf P_{M,{\rm m}},\mathbf J_{M,{\rm m}},\mathbf H_{M,{\rm m}})
    \longmapsto
    (-\mathbf P_E,\mathbf J_E,-\mathbf H_E).
\]
Therefore the Harrison commutation relations are preserved, and the sectorial Casimirs are interchanged:
\[
    \mathcal C_{E}^{\rm m}
    =
    \mathcal C_M,
    \qquad
    \mathcal C_{M}^{\rm m}
    =
    \mathcal C_E.
\]
The signs in \(\mathbf P\) and \(\mathbf H\) do not affect the quadratic Casimirs.

At the Hamiltonian level, the geodesic Hamiltonian is invariant, i.e. 
\(
    H_{\rm M}^{\rm m}=H_{\rm M}.
\)
Thus the LWP magnetic map is a canonical target-space duality. It preserves the geodesic Hamiltonian and therefore preserves the \(k\)-quadrature, while it exchanges the electric and magnetic visible Noether charges.

\subsection{Frozen ModMax under the magnetic LWP map}

Under the magnetic LWP map, the frozen ModMax coefficients transform as
\[
    w_{\rm m}
    =
    \frac{w_0}{w_0^2+v_0^2},
    \qquad
    v_{\rm m}
    =
    -\frac{v_0}{w_0^2+v_0^2}.
\]
Together with
\[
    p_{\psi_{\rm m}}=p_\chi+\psi p_\epsilon,
    \qquad
    p_{\chi_{\rm m}}+\psi_{\rm m}p_{\epsilon_{\rm m}}=-p_\psi,
\]
this gives
\(
    H_{\rm MM}^{\rm m}=H_{\rm MM}.
\)
Hence the magnetic LWP map remains canonical in the frozen ModMax target, provided the frozen coefficients transform as above.

The Noether charges transform precisely as
\(
    \omega_{0,{\rm m}}=\omega_0,
    \quad
    \mathtt p_{\rm m}=-\mathtt q,
    \quad
    \mathtt q_{\rm m}=\mathtt p.
\)
Consequently, the twist quadrature remains the same, whereas the electromagnetic dual-harmonic quadrature preserves its structural form, provided one substitutes the Maxwell magnetic combination with the frozen ModMax combination given above.

\section{Two-harmonic scalar seeds and sectorial frozen-ModMax transformations}
\label{sec:two-harmonic-frozen-modmax-transformations}

We now implement the sectorial maps \eqref{eq:Mapeos-HarrisonElectrico-MM},
\eqref{eq:Mapeos-HarrisonMagnetico-MM}, and \eqref{eq:Mapeos-Ehlers-M} on a scalar–vacuum Weyl seed characterized by two independent harmonic functions. In terms of the generalized Ernst variables, this is the direct counterpart of the scalar multipolar Weyl construction, wherein the gravitational and scalar sectors are governed by distinct harmonic functions \cite{Herdeiro:2024oxn}. 

\subsection{Two-harmonic scalar seed}

Let \(\mathfrak s_f(\rho,z)\) and \(\mathfrak s_\kappa(\rho,z)\) be two
independent Weyl-harmonic functions, using the notation given in \cite{Bixano:2026ouq}
\[
    \rho^{-1}D(\rho D\mathfrak s_f)=0,
    \qquad
    \rho^{-1}D(\rho D\mathfrak s_\kappa)=0.
\]
We take the static neutral seed
\begin{equation}\label{semilla}
        f_0=f_\star e^{\lambda\mathfrak s_f},
    \qquad
    \kappa_0=\kappa_\star e^{\tau \mathfrak s_\kappa},
    \qquad
    \epsilon_0^{\rm twist}=\psi_0=\chi_0=0.
\end{equation}
Here \(\epsilon_0^{\rm twist}\) denotes the vanishing twist potential of the seed, not the constant \(\epsilon_0=\pm1\) and \(\lambda,\tau\) are integration constants. Equivalently,
\[
    \ln f_0=\ln f_\star+\lambda\mathfrak s_f,
    \qquad
    \ln\kappa_0=\ln\kappa_\star+\tau \mathfrak s_\kappa .
\]
Thus \(\ln f_0\) and \(\ln\kappa_0\) are independently harmonic.

Since
\(
    \kappa^2=e^{-2\alpha_0\phi},
\)
the scalar field is
\[
    \phi_0
    =
    -\frac{1}{\alpha_0}\ln\kappa_\star
    -
    \frac{\tau}{\alpha_0}\mathfrak s_\kappa .
\]
For this seed,
\[
    Df_0=\lambda f_0D\mathfrak s_f,
    \qquad
    D\kappa_0=\tau\kappa_0D\mathfrak s_\kappa .
\]
Employing the definitions in \eqref{eq:1Formas-ABC}, as introduced in \cite{Bixano:2026xum,Bixano:2026ouq},
\[
    A_0=\frac{\lambda}{2}D\mathfrak s_f,\qquad
    B_0=0,\qquad
    C_0=-\tau D\mathfrak s_\kappa .
\]
This directly solves the \(ABC-\)system, equations (15) in \cite{Bixano:2026xum} and (20) in \cite{Bixano:2026ouq} consequently. Since \(A_0=\bar A_0\) and \(B_0=0\), the \(A\)-equation reduces to
\(
    \rho^{-1}D(\rho A_0)=0,
\)
which follows from the harmonicity of \(\mathfrak s_f\). Similarly, the
\(C\)-equation reduces to
\(
    \rho^{-1}D(\rho C_0)=0,
\)
which is a direct consequence of the harmonicity of \(\mathfrak s_\kappa\). Consequently, the two-harmonic seed constitutes an exact scalar-vacuum seed of the generalized Ernst system discussed in \cite{Herdeiro:2024oxn}.

The corresponding \(k\) function is obtained by two independent quadratures:
\[
    k_0
    =
    k_\star
    +
    \frac{\lambda^2}{4}\mathfrak R_f
    +
    \frac{\epsilon_0}{\alpha_0^2}\tau^2\mathfrak R_\kappa ,
\]
where
\(
    (\mathfrak R_f)_{,\zeta}
    =
    \rho\,\mathfrak s_{f,\zeta}^{\,2},
    \qquad
    (\mathfrak R_\kappa)_{,\zeta}
    =
    \rho\,\mathfrak s_{\kappa,\zeta}^{\,2}
\), and the energies of the geodesics are \(\frac{\lambda^2}{4}\mathfrak,\frac{\epsilon_0}{\alpha_0^2}\tau^2 \) respectively for \(s_{f},s_{\kappa}\).

\subsection{Electric frozen-ModMax Harrison branch}

We first apply the electric frozen-ModMax Harrison map
\eqref{eq:Mapeos-HarrisonElectrico-MM}. Since the seed has
\(\psi_0=0\), one obtains
\[
    \Delta_E^{\rm MM}
    =
    1-
    \frac{\mathtt q^2}{4\gamma}
    \frac{f_0}{\Lambda_E\kappa_0^2}.
\]
Using the two-harmonic seed, this becomes
\[
    \Delta_E^{\rm MM}
    =
    1-
    \frac{\mathtt q^2}{4\gamma\Lambda_E}
    \frac{f_\star}{\kappa_\star^2}
    \exp\!\left(
        \lambda\mathfrak s_f-2\tau\mathfrak s_\kappa
    \right).
\]
The electric branch is therefore
\begin{equation}\label{eq:Solucion2Arm-EH}
    \begin{aligned}
        &f_E
        =
        f_0\left(\Delta_E^{\rm MM}\right)^{-1/(2\gamma)},
        \qquad
        \kappa_E
        =
        \kappa_0\left(\Delta_E^{\rm MM}\right)^{\beta/(2\gamma)},
        \\
        &\psi_E
    =
    \frac{
        \dfrac{\mathtt q}{2\gamma}
        \dfrac{f_0}{\Lambda_E\kappa_0^2}
    }{
        \Delta_E^{\rm MM}
    },
    \qquad
    \epsilon_E=0,
    \qquad
    \chi_E=0.
    \end{aligned}
\end{equation}

Thus the electric transformation is controlled by the combination
\[
    \frac{f_0}{\kappa_0^2}
    =
    \frac{f_\star}{\kappa_\star^2}
    \exp\!\left(
        \lambda\mathfrak s_f-2\tau\mathfrak s_\kappa
    \right).
\]

\subsection{Magnetic frozen-ModMax Harrison branch}

Now we apply the magnetic frozen-ModMax Harrison map
\eqref{eq:Mapeos-HarrisonMagnetico-MM}. Since the seed has
\(\chi_0=0\), one obtains
\[
    \Delta_M^{\rm MM}
    =
    1-
    \frac{\mathtt p^2}{4\gamma}
    w_0f_0\kappa_0^2.
\]
Explicitly,
\[
    \Delta_M^{\rm MM}
    =
    1-
    \frac{\mathtt p^2w_0}{4\gamma}
    f_\star\kappa_\star^2
    \exp\!\left(
        \lambda\mathfrak s_f+2\tau\mathfrak s_\kappa
    \right).
\]
The magnetic branch is
\begin{equation}\label{eq:Solucion2Arm-MH}
    \begin{aligned}
        &f_M
        =
        f_0\left(\Delta_M^{\rm MM}\right)^{-1/(2\gamma)},
        \qquad
        \kappa_M
        =
        \kappa_0\left(\Delta_M^{\rm MM}\right)^{-\beta/(2\gamma)},
        \\
        &\chi_M
        =
        \frac{
            \dfrac{\mathtt p}{2\gamma}
            w_0f_0\kappa_0^2
        }{
            \Delta_M^{\rm MM}
        },
        \qquad
        \epsilon_M=0,
        \qquad
        \psi_M=0.
    \end{aligned}
\end{equation}

Thus the magnetic transformation is controlled by the dual combination
\[
    f_0\kappa_0^2
    =
    f_\star\kappa_\star^2
    \exp\!\left(
        \lambda\mathfrak s_f+2\tau\mathfrak s_\kappa
    \right).
\]

\subsection{Ehlers branch}

Finally, we apply the sectorial Ehlers map
\eqref{eq:Mapeos-Ehlers-M}. For the seed
\[
    \epsilon_0^{\rm twist}=0,
    \qquad
    f=f_0,
    \qquad
    \kappa=\kappa_0,
\]
one obtains
\begin{equation}\label{eq:Solucion2Arm-Ehlers}
    \begin{aligned}
        &f_{\rm Eh}
        =
        \frac{f_0}{1+s^2f_0^2},
        \qquad
        \kappa_{\rm Eh}=\kappa_0,
        \\
        &\epsilon_{\rm Eh}
    =
    -\frac{s f_0^2}{1+s^2f_0^2}.
    \end{aligned}
\end{equation}
The electromagnetic potentials remain zero,
\(
    \psi_{\rm Eh}=0,
    \qquad
    \chi_{\rm Eh}=0.
\)
Thus, the Ehlers branch generates a gravitational twist, leaving the scalar sector unchanged and the electromagnetic sector unexcited.

\subsection{Final quadratures after the sectorial maps}

\subsubsection*{Electric frozen-ModMax branch}
Taking into account the solution \eqref{eq:Solucion2Arm-EH}, one has
\(
    D\ln\Delta_E^{\rm MM}
    =
    \frac{\Delta_E^{\rm MM}-1}{\Delta_E^{\rm MM}}
    \left(
        \lambda D\mathfrak s_f
        -
        2\tau D\mathfrak s_\kappa
    \right).
\)
Thus, one froms \eqref{eq:1Formas-ABC} take the form
\[
    A_E
    =
    \left[
        \frac{\lambda}{2}
        -
        \frac{\lambda}{4\gamma}
        \frac{\Delta_E^{\rm MM}-1}{\Delta_E^{\rm MM}}
    \right]D\mathfrak s_f
    +
    \left[
        \frac{\tau}{2\gamma}
        \frac{\Delta_E^{\rm MM}-1}{\Delta_E^{\rm MM}}
    \right]D\mathfrak s_\kappa,
\]
and
\[
    C_E
    =
    -
    \left[
        \frac{\beta\lambda}{2\gamma}
        \frac{\Delta_E^{\rm MM}-1}{\Delta_E^{\rm MM}}
    \right]D\mathfrak s_f
    +
    \left[
        -\tau
        +
        \frac{\beta\tau}{\gamma}
        \frac{\Delta_E^{\rm MM}-1}{\Delta_E^{\rm MM}}
    \right]D\mathfrak s_\kappa .
\]
Moreover, the electric contribution is
\[
    B_E\bar B_E
    =
    \frac{1-\Delta_E^{\rm MM}}
    {4\gamma\left(\Delta_E^{\rm MM}\right)^2}
    \left(
        \lambda D\mathfrak s_f
        -
        2\tau D\mathfrak s_\kappa
    \right)^2 .
\]
Substituting these three expressions in
\(
    A_E\bar A_E
    -
    B_E\bar B_E
    +
    \frac{\epsilon_0}{\alpha_0^2}C_E^2,
\)
and using
\(
    \beta=\frac{\alpha_0^2}{2\epsilon_0},
    \qquad
    4\gamma=2\beta+1,
\)
all terms depending on \(\Delta_E^{\rm MM}\) cancel. The final result is
\[
    A_E\bar A_E
    -
    B_E\bar B_E
    +
    \frac{\epsilon_0}{\alpha_0^2}C_E^2
    =
    \frac{\lambda^2}{4}
    (D\mathfrak s_f)^2
    +
    \frac{\epsilon_0}{\alpha_0^2}\tau^2
    (D\mathfrak s_\kappa)^2 .
\]
Therefore the electric branch has final \(k\)-quadrature coefficients
\[
    \mathcal K^{E}_{f}
    =
    \frac{\lambda^2}{4},
    \qquad
    \mathcal K^{E}_{f\kappa}
    =
    0,
    \qquad
    \mathcal K^{E}_{\kappa}
    =
    \frac{\epsilon_0}{\alpha_0^2}\tau^2 .
\]
Thus the electric Harrison map changes \(f,\kappa,\psi\), but it does not change the coefficients entering the \(k\)-quadrature, and the other quadratures are
\[
    A_t^{(E)}=\frac12\psi_E,
    \qquad
    A_\varphi^{(E)}=0,
    \qquad
    \omega_E=0 .
\]

\subsubsection*{Magnetic frozen-ModMax branch}

For the magnetic configuration given in \eqref{eq:Solucion2Arm-MH}, we obtain
\[
    D\ln\Delta_M^{\rm MM}
    =
    \frac{\Delta_M^{\rm MM}-1}{\Delta_M^{\rm MM}}
    \left(
        \lambda D\mathfrak s_f
        +
        2\tau D\mathfrak s_\kappa
    \right).
\]
From the final transformed potentials,
\[
    A_M
    =
    \left[
        \frac{\lambda}{2}
        -
        \frac{\lambda}{4\gamma}
        \frac{\Delta_M^{\rm MM}-1}{\Delta_M^{\rm MM}}
    \right]D\mathfrak s_f
    -
    \left[
        \frac{\tau}{2\gamma}
        \frac{\Delta_M^{\rm MM}-1}{\Delta_M^{\rm MM}}
    \right]D\mathfrak s_\kappa,
\]
and
\[
    C_M
    =
    \left[
        \frac{\beta\lambda}{2\gamma}
        \frac{\Delta_M^{\rm MM}-1}{\Delta_M^{\rm MM}}
    \right]D\mathfrak s_f
    +
    \left[
        -\tau
        +
        \frac{\beta\tau}{\gamma}
        \frac{\Delta_M^{\rm MM}-1}{\Delta_M^{\rm MM}}
    \right]D\mathfrak s_\kappa .
\]
The magnetic contribution is
\[
    B_M\bar B_M
    =
    \frac{1-\Delta_M^{\rm MM}}
    {4\gamma\left(\Delta_M^{\rm MM}\right)^2}
    \left(
        \lambda D\mathfrak s_f
        +
        2\tau D\mathfrak s_\kappa
    \right)^2 .
\]
Substituting in
\(
    A_M\bar A_M
    -
    B_M\bar B_M
    +
    \frac{\epsilon_0}{\alpha_0^2}C_M^2
\)
again cancels all \(\Delta_M^{\rm MM}\)-dependent terms. Therefore
\[
    A_M\bar A_M
    -
    B_M\bar B_M
    +
    \frac{\epsilon_0}{\alpha_0^2}C_M^2
    =
    \frac{\lambda^2}{4}
    (D\mathfrak s_f)^2
    +
    \frac{\epsilon_0}{\alpha_0^2}\tau^2
    (D\mathfrak s_\kappa)^2 .
\]
Thus the magnetic branch has final coefficients
\[
    \mathcal K^{M}_{f}
    =
    \frac{\lambda^2}{4},
    \qquad
    \mathcal K^{M}_{f\kappa}
    =
    0,
    \qquad
    \mathcal K^{M}_{\kappa}
    =
    \frac{\epsilon_0}{\alpha_0^2}\tau^2 .
\]

The branch is purely magnetic:
\[
    A_t^{(M)}=0,
    \qquad
    \omega_M=0 ,
\]
\[
    A_\varphi^{(M)}
    =
    -
    \frac{\mathtt p}{4\gamma}
    \left(
        \lambda\widetilde{\mathfrak s}_f
        +
        2\tau\widetilde{\mathfrak s}_\kappa
    \right).
\]

\subsubsection*{Ehlers branch}

For the Ehlers map \eqref{eq:Solucion2Arm-Ehlers},  one obtains
\(
    A_{\rm Eh}
    =
    \frac{1}{2f_{\rm Eh}}
    \left(
        Df_{\rm Eh}-iD\epsilon_{\rm Eh}
    \right),
    \qquad
    C_{\rm Eh}=-D\ln\kappa_0.
\)
A direct calculation gives
\(
    A_{\rm Eh}\bar A_{\rm Eh}
    =
    \frac{\lambda^2}{4}
    (D\mathfrak s_f)^2,
    \qquad
    C_{\rm Eh}
    =
    -\tau D\mathfrak s_\kappa.
\)
Since \(B_{\rm Eh}=0\), we get
\(
    A_{\rm Eh}\bar A_{\rm Eh}
    -
    B_{\rm Eh}\bar B_{\rm Eh}
    +
    \frac{\epsilon_0}{\alpha_0^2}C_{\rm Eh}^2
    =
    \frac{\lambda^2}{4}
    (D\mathfrak s_f)^2
    +
    \frac{\epsilon_0}{\alpha_0^2}\tau^2
    (D\mathfrak s_\kappa)^2 .
\)
Therefore
\[
    \mathcal K^{\rm Eh}_{f}
    =
    \frac{\lambda^2}{4},
    \qquad
    \mathcal K^{\rm Eh}_{f\kappa}
    =
    0,
    \qquad
    \mathcal K^{\rm Eh}_{\kappa}
    =
    \frac{\epsilon_0}{\alpha_0^2}\tau^2 
\]
and
\[
    \omega_{\rm Eh}
    =
    -
    2s\lambda\,\widetilde{\mathfrak s}_f,
    \quad 
    A_t^{({\rm Eh})}=0,
    \quad
    A_\varphi^{({\rm Eh})}=0,
\]
i.e. the only dual harmonic function operating on \(\omega\) is the one associated with the gravitational potential \(\mathfrak s_f\).

\section{Conclusions}   %

The visible transformations form a solvable algebra of the form
\(
    \mathfrak h_3\rtimes(\mathbb R\oplus\mathbb R),
\)
where the Heisenberg factor is generated by the translation-type symmetries. By contrast, the Ehlers and Harrison transformations are sectorial: the Ehlers map acts on the gravito-rotational block \( \psi=\chi=0\), while the electric and magnetic Harrison maps act on the corresponding static electromagnetic blocks \(\epsilon=\chi=0\) or \(\epsilon=\psi=0\). This distinction is essential. For generic scalar coupling, the electric and magnetic Harrison maps are not global isometries of the complete five-dimensional target space.

In the frozen sector the ModMax coefficients are constant, so the electric and magnetic Harrison maps remain as explicit sectorial transformations. The coexistence of both Harrison sectors selects this regime, since the compatibility conditions require constant ModMax coefficients. Frozen ModMax does not destroy the sectorial maps, it only renormalizes the electric and magnetic blocks. Thus, the frozen EMMSF theory preserves the sectorial Harrison structure, if and only if \(v,w\) are constant.

A central result is the Hamiltonian interpretation of the metric quadratures. In a pure harmonic branch, the target-space curve is an affinely parametrized geodesic whose conserved geodesic Hamiltonian becomes the constant coefficient in the quadrature of the Weyl function \(k\). In the \(A,B,C\) description, if
\(
    A=a(\mathfrak s)D\mathfrak s,\quad
    B=b(\mathfrak s)D\mathfrak s,\quad
    C=c(\mathfrak s)D\mathfrak s,
\)
the conserved geodesic energy is
\(
    \mathcal K_0
    =
    a\bar a-b\bar b
    +
    \frac{\epsilon_0}{\alpha_0^2}c^2 .
\)
Thus the coefficient in the \(k\)-quadrature is precisely the Hamiltonian of the affine target-space curve, so integrating \(k\) is directly governed by the target-space energy.

The remaining physical quadratures also have a direct Noether interpretation. The dragging potential \(\omega\) is fixed by the Noether charge along the twist direction \(\epsilon\), and the azimuthal electromagnetic potential \(A_\varphi\) by the Noether charge along the magnetic potential direction \(\chi\). In harmonic branches these charges integrate through the dual harmonic function \(\widetilde{\mathfrak s}\). Hence the three physical functions
\(
    k,\quad \omega,\quad A_\varphi
\) given by \eqref{eq:k-quadrature-harmonic},\eqref{eq:omega-harmonic-quadrature} and \eqref{eq:Aphi-harmonic-quadrature} are determined by three target-space quantities: the Hamiltonian and the Noether charges in the \(\epsilon\) and \(\chi\) directions, which is a key conceptual outcome of the construction.

The quantities
\(
    a\bar a,  b\bar b, c^2
\)
measure the gravito-rotational, electromagnetic, and scalar contributions to the target-space kinetic energy.The Harrison subsectors are different. Their Casimirs are not given only by the electric or magnetic pieces of \(b\bar b\). Rather, each Harrison map couples an electromagnetic potential to a specific gravito-scalar combination. This is already visible in the finite maps: the electric Harrison transformation depends on \(f/\kappa^2\), whereas the magnetic
Harrison transformation depends on \(f\kappa^2\). Therefore the electric and magnetic contributions
\(
    b\bar b=b_E+b_M
\)
are only the electromagnetic parts of the corresponding Harrison Casimirs. The full sectorial Casimirs also include the gravito-scalar directions selected by the Harrison maps. In this sense, the Casimir structure gives a geometric decomposition of the Hamiltonian coefficient that controls the \(k\)-quadrature, rather than being only an abstract
algebraic invariant. In the frozen ModMax case the same statement holds with the constant renormalizations
\[
    \frac{f}{\kappa^2}
    \longmapsto
    \frac{f}{\Lambda_E\kappa^2},
    \qquad
    f\kappa^2
    \longmapsto
    w_0f\kappa^2 .
\]
We also clarified the role of the electric-magnetic Lewis-Weyl-Papapetrou map. This map is not a Harrison transformation, but a discrete frame-duality exchanging the temporal and axial reductions. In the scalar-coupled theory it acts nontrivially on the scalar sector, mapping \(\kappa\) to \(\kappa^{-1}\) and therefore reversing the scalar profile. Hence, when \(D\kappa\neq0\), the magnetic LWP frame generally defines a four-dimensional configuration that is not physically equivalent to the original electric-frame one, even if the reduced functions are formally related.

Finally, we applied the frozen-ModMax sectorial maps \eqref{eq:Mapeos-HarrisonElectrico-MM},
\eqref{eq:Mapeos-HarrisonMagnetico-MM}, and \eqref{eq:Mapeos-Ehlers-M} to two-harmonic scalar-vacuum Weyl seed \eqref{semilla}. The harmonic \(\mathfrak s_f\) controls \(\ln f_0\), while \(\mathfrak s_\kappa\) controls \(\ln\kappa_0\). The electric and magnetic branches couple to the scalar harmonic with opposite weights:
\[
    \frac{f_0}{\Lambda_E\kappa_0^2}
    \propto
    e^{\lambda\mathfrak s_f-2\tau\mathfrak s_\kappa},
    \qquad
    w_0f_0\kappa_0^2
    \propto
    e^{\lambda\mathfrak s_f+2\tau\mathfrak s_\kappa}.
\]
Thus the scalar harmonic screens the electric frozen-ModMax branch and enhances the magnetic one, or vice versa depending on the sign of \(\tau\mathfrak s_\kappa\). The constants \(\Lambda_E\) and \(w_0\) encode the frozen-ModMax renormalization of the electric and magnetic blocks.

The magnetic potential \(A_\varphi^{(M)}\) depends on the dual harmonics of both \(\mathfrak s_f\) and \(\mathfrak s_\kappa\), whereas the Ehlers dragging potential \(\omega_{\rm Eh}\) depends only on the dual harmonic of \(\mathfrak s_f\). In all three branches the \(k\)-quadrature keeps the same two energy contributions,
\[
    \frac{\lambda^2}{4},
    \qquad
    \frac{\epsilon_0}{\alpha_0^2}\tau^2,
\]
associated with the gravitational and scalar harmonics, respectively. Therefore a single two-harmonic scalar seed generates three exact sectorial frozen-ModMax orbits: electric, magnetic, and Ehlers-twisted, all with explicit quadratures for \(k\), \(\omega\), and \(A_\varphi\).

\section{Acknowledgements}
The author would like to thank \emph{Ana Bokulić} and \emph{Carlos A. R. Herdeiro} for valuable discussions and insightful comments which motivated part of the analysis presented in the section \ref{Sec:AnaBokulic} of this work. 

LB thanks SECIHTI-M\'exico for the doctoral grant.
This work was also partially supported by SECIHTI M\'exico under grants SECIHTI CBF-2025-G-1720 and CBF-2025-G-176. 


\begin{appendices}
\section{Coordinates protection}\label{Apendix:Coordinates protection}
 
\subsection{Potential differential equations}

\end{appendices}
\bibliographystyle{elsarticle-harv} 
\bibliography{Bibliografia}

@article{MATOS1988423,
title = {SL(3, R) representation for invariance transformations in five-dimensional gravity},
journal = {Physics Letters A},
volume = {131},
number = {7},
pages = {423-426},
year = {1988},
issn = {0375-9601},
doi = {https://doi.org/10.1016/0375-9601(88)90292-7},
author = {Tonatiuh Matos}
}

@article{Matos:1986,
    author = "Matos, T.",
    title = "{The linear problem for the five-dimensional projective field theory}",
    doi = "10.1002/asna.2113070521",
    journal = "Astron. Nachr.",
    volume = "307",
    pages = "317-320",
    year = "1986"
}

@article{Matos:2010pcd,
    author = "Matos, Tonatiuh",
    title = "{Class of Einstein-Maxwell Phantom Fields: Rotating and Magnetised Wormholes}",
    eprint = "0902.4439",
    archivePrefix = "arXiv",
    primaryClass = "gr-qc",
    reportNumber = "CIEA-09-GR5",
    doi = "10.1007/s10714-010-0976-6",
    journal = "Gen. Rel. Grav.",
    volume = "42",
    pages = "1969--1990",
    year = "2010"
}

@article{Matos:2000ai,
    author = "Matos, Tonatiuh and Nunez, Dario and Estevez, Gabino and Rios, Maribel",
    title = "{Rotating 5-D Kaluza-Klein space-times from invariant transformations}",
    eprint = "gr-qc/0001039",
    archivePrefix = "arXiv",
    reportNumber = "CINVESTAV-00-FIS-8",
    doi = "10.1023/A:1001982001694",
    journal = "Gen. Rel. Grav.",
    volume = "32",
    pages = "1499--1525",
    year = "2000"
}

@article{Matos:2000za,
    author = "Matos, Tonatiuh and Nunez, Dario and Rios, Maribel",
    title = "{Class of Einstein-Maxwell dilatons, an ansatz for new families of rotating solutions}",
    eprint = "gr-qc/0008068",
    archivePrefix = "arXiv",
    doi = "10.1088/0264-9381/17/18/323",
    journal = "Class. Quant. Grav.",
    volume = "17",
    pages = "3917--3934",
    year = "2000"
}

@article{Matos:2009rp,
    author = "Matos, Tonatiuh and Miranda, Galaxia and Sanchez-Sanchez, Ruben and Wiederhold, Petra",
    title = "{Class of Einstein-Maxwell-Dilaton-Axion Space-Times}",
    eprint = "0905.4097",
    archivePrefix = "arXiv",
    primaryClass = "gr-qc",
    reportNumber = "CIEA-09-GR14",
    doi = "10.1103/PhysRevD.79.124016",
    journal = "Phys. Rev. D",
    volume = "79",
    pages = "124016",
    year = "2009"
}

@article{Ernst:1967wx,
    author = "Ernst, Frederick J.",
    title = "{New formulation of the axially symmetric gravitational field problem}",
    doi = "10.1103/PhysRev.167.1175",
    journal = "Phys. Rev.",
    volume = "167",
    pages = "1175--1179",
    year = "1968"
}

@article{Ernst:1967by,
    author = "Ernst, Frederick J.",
    title = "{New Formulation of the Axially Symmetric Gravitational Field Problem. II}",
    doi = "10.1103/PhysRev.168.1415",
    journal = "Phys. Rev.",
    volume = "168",
    pages = "1415--1417",
    year = "1968"
}

@article{Matos:1994hm,
    author = "Matos, Tonatiuh",
    title = "{5-D axisymmetric stationary solutions as harmonic maps}",
    eprint = "gr-qc/9401009",
    archivePrefix = "arXiv",
    doi = "10.1063/1.530590",
    journal = "J. Math. Phys.",
    volume = "35",
    pages = "1302--1321",
    year = "1994"
}

@article{Bokulic:2025ucy,
    author = "Bokuli{\'c}, Ana and Herdeiro, Carlos A. R.",
    title = "{Generalised Harrison transformations and black diholes in Einstein-ModMax}",
    eprint = "2507.16926",
    archivePrefix = "arXiv",
    primaryClass = "gr-qc",
    doi = "10.1007/JHEP10(2025)091",
    journal = "JHEP",
    volume = "10",
    pages = "091",
    year = "2025"
}

@article{Bixano:2026ouq,
    author = "Bixano, Leonel and Matos, Tonatiuh",
    title = "{Generalized Einstein-ModMax-ScalarField theories and new exact solutions}",
    eprint = "2603.26073",
    archivePrefix = "arXiv",
    primaryClass = "gr-qc",
    month = "3",
    year = "2026"
}

@article{Bixano:2026xum,
    author = "Bixano, Leonel and Matos, Tonatiuh",
    title = "{Generalized Ernst Potentials for arbitrary Dilatonic Theories}",
    eprint = "2603.02384",
    archivePrefix = "arXiv",
    primaryClass = "gr-qc",
    month = "3",
    year = "2026"
}

@article{Harrison:1968wue,
    author = "Harrison, B. Kent",
    title = "{New Solutions of the Einstein-Maxwell Equations from Old}",
    doi = "10.1063/1.1664508",
    journal = "J. Math. Phys.",
    volume = "9",
    number = "11",
    pages = "1744",
    year = "1968"
}

@article{Kinnersley:1977pg,
    author = "Kinnersley, W.",
    title = "{Symmetries of the Stationary Einstein-Maxwell Field Equations. 1.}",
    doi = "10.1063/1.523458",
    journal = "J. Math. Phys.",
    volume = "18",
    pages = "1529--1537",
    year = "1977"
}

@article{Galtsov:1994pd,
    author = "Galtsov, D. V. and Kechkin, O. V.",
    title = "{Ehlers-Harrison type transformations in dilaton - axion gravity}",
    eprint = "hep-th/9407155",
    archivePrefix = "arXiv",
    reportNumber = "MSU-DTP-94-2-REV, MSU-DTP-94-2",
    doi = "10.1103/PhysRevD.50.7394",
    journal = "Phys. Rev. D",
    volume = "50",
    pages = "7394--7399",
    year = "1994"
}

@article{Herrera-Aguilar:1998oct,
    author = "Herrera-Aguilar, Alfredo and Kechkin, Oleg",
    title = "{Charging symmetries and linearizing potentials for Einstein-Maxwell dilaton - axion theory}",
    eprint = "hep-th/9806247",
    archivePrefix = "arXiv",
    doi = "10.1142/S0217732398002011",
    journal = "Mod. Phys. Lett. A",
    volume = "13",
    pages = "1907--1914",
    year = "1998"
}

@article{Herdeiro:2024oxn,
    author = "Herdeiro, Carlos A. R.",
    title = "{Black holes in scalar multipolar universes}",
    eprint = "2410.12950",
    archivePrefix = "arXiv",
    primaryClass = "gr-qc",
    doi = "10.1016/j.physletb.2024.139160",
    journal = "Phys. Lett. B",
    volume = "860",
    pages = "139160",
    year = "2025"
}

@article{Galtsov:1995mb,
    author = "Galtsov, D. V. and Garcia, A. A. and Kechkin, O. V.",
    title = "{Symmetries of the stationary Einstein-Maxwell dilaton theory}",
    eprint = "hep-th/9504155",
    archivePrefix = "arXiv",
    reportNumber = "CINVESTAV-GRG-94-12-REV, CINVESTAV-GRG-94-12",
    doi = "10.1088/0264-9381/12/12/007",
    journal = "Class. Quant. Grav.",
    volume = "12",
    pages = "2887--2903",
    year = "1995"
}

@inproceedings{Galtsov:1995ssj,
    author = "Gal'tsov, D. V. and Kechkin, O. V.",
    title = "{Hidden symmetries in dilaton - axion gravity}",
    booktitle = "{International Workshop on Geometry and Integrable Models}",
    eprint = "gr-qc/9606014",
    archivePrefix = "arXiv",
    reportNumber = "DTP-MSU-95-26",
    pages = "78--95",
    month = "10",
    year = "1995"
}

@article{Galtsov:1996qko,
    author = "Gal'tsov, D. V. and Letelier, P. S.",
    title = "{Ehlers-Harrison transformations and black holes in dilaton - axion gravity with multiple vector fields}",
    eprint = "gr-qc/9612007",
    archivePrefix = "arXiv",
    doi = "10.1103/PhysRevD.55.3580",
    journal = "Phys. Rev. D",
    volume = "55",
    pages = "3580--3592",
    year = "1997"
}

@article{Galtsov:1994sjr,
    author = "Gal'tsov, D. V.",
    title = "{Integrable systems in stringy gravity}",
    eprint = "hep-th/9410217",
    archivePrefix = "arXiv",
    reportNumber = "MSU-DTP-94-21",
    doi = "10.1103/PhysRevLett.74.2863",
    journal = "Phys. Rev. Lett.",
    volume = "74",
    pages = "2863--2866",
    year = "1995"
}

@article{Manko:2019gvj,
    author = "Manko, V. S. and Mej{\'\i}a, I. M. and Ruiz, E.",
    title = "{Metric of a rotating charged magnetized sphere}",
    eprint = "1912.08884",
    archivePrefix = "arXiv",
    primaryClass = "gr-qc",
    doi = "10.1016/j.physletb.2020.135286",
    journal = "Phys. Lett. B",
    volume = "803",
    pages = "135286",
    year = "2020"
}

@article{Matos:1995my,
    author = "Matos, Tonatiuh and Nunez, Dario and Quevedo, Hernando",
    title = "{Class of Einstein-Maxwell dilatons}",
    eprint = "gr-qc/9510042",
    archivePrefix = "arXiv",
    doi = "10.1103/PhysRevD.51.R310",
    journal = "Phys. Rev. D",
    volume = "51",
    pages = "R310--R313",
    year = "1995"
}

@article{Matos:1994qv,
    author = "Matos, T. and Plebanski, J.",
    title = "{Axisymmetric stationary solutions as harmonic maps}",
    eprint = "gr-qc/9402044",
    archivePrefix = "arXiv",
    reportNumber = "CINVESTAV-12-93",
    doi = "10.1007/BF02108050",
    journal = "Gen. Rel. Grav.",
    volume = "26",
    pages = "477",
    year = "1994"
}

@article{Matos:1996km,
    author = "Matos, Tonatiuh and Mora, Cesar",
    title = "{Stationary dilatons with arbitrary electromagnetic field}",
    eprint = "hep-th/9610013",
    archivePrefix = "arXiv",
    reportNumber = "CINVESTAV-FIS-15-96, UMSNH-IFM-12-96",
    doi = "10.1088/0264-9381/14/8/027",
    journal = "Class. Quant. Grav.",
    volume = "14",
    pages = "2331--2340",
    year = "1997"
}

@article{Barrientos:2023dlf,
    author = "Barrientos, Jos{\'e} and Cisterna, Adolfo and Pallikaris, Konstantinos",
    title = "{Pleban{\'s}ki{\textendash}Demia{\'n}ski {\`a} la Ehlers{\textendash}Harrison: exact rotating and accelerating type I black holes}",
    eprint = "2309.13656",
    archivePrefix = "arXiv",
    primaryClass = "gr-qc",
    doi = "10.1007/s10714-024-03304-x",
    journal = "Gen. Rel. Grav.",
    volume = "56",
    number = "9",
    pages = "111",
    year = "2024"
}

@article{Kinnersley:1977ph,
    author = "Kinnersley, W. and Chitre, D. M.",
    title = "{Symmetries of the Stationary Einstein-Maxwell Field Equations. 2.}",
    doi = "10.1063/1.523459",
    journal = "J. Math. Phys.",
    volume = "18",
    pages = "1538--1542",
    year = "1977"
}

@article{Kinnersley:1978pz,
    author = "Kinnersley, W. and Chitre, D. M.",
    title = "{Symmetries of the Stationary Einstein-Maxwell Field Equations. 3.}",
    doi = "10.1063/1.523912",
    journal = "J. Math. Phys.",
    volume = "19",
    pages = "1926--1931",
    year = "1978"
}

@book{Knapp:1996,
  author    = {Knapp, Anthony W.},
  title     = {Lie Groups Beyond an Introduction},
  series    = {Progress in Mathematics},
  volume    = {140},
  edition   = {1},
  publisher = {Birkh{\"a}user Boston},
  address   = {Boston, MA},
  year      = {1996},
  doi       = {10.1007/978-1-4757-2453-0},
  isbn      = {978-1-4757-2453-0}
}

\end{document}